%% file: VH.tex
\documentclass[a4paper,11pt]{article}

\usepackage{setup}
\usepackage{jheppub}

\usepackage{cite}
\usepackage{tikz}
\usepackage{ifthen}
\usepackage{subcaption}
\usepackage{siunitx}
\usepackage{enumitem}
\usepackage{booktabs}

\DeclareSIUnit\fb{\femto\barn}

\renewcommand{\tabcolsep}{1em}

\usetikzlibrary{shapes,snakes,arrows.meta,calc}
\usetikzlibrary{shapes.callouts}
\usetikzlibrary{decorations.pathmorphing}
\usetikzlibrary{decorations.markings}
\usetikzlibrary{backgrounds}
\usetikzlibrary{positioning}

\definecolor{lightgray}{HTML}{A6A39A}
\definecolor{darkgray}{HTML}{504E48}
\definecolor{silver}{HTML}{E0DFDE}
\definecolor{brown}{HTML}{5F4541}
\definecolor{beige}{HTML}{DCCCAC}
\definecolor{green}{HTML}{345F53}
\definecolor{yellow}{HTML}{F6B65A}
\definecolor{blue}{HTML}{568BCF}
\definecolor{red}{HTML}{AE1932}
\definecolor{orange}{HTML}{D16F15}

\makeatletter
\newcommand{\myitem}[1]{%
	\item[#1]\protected@edef\@currentlabel{#1}%
}
\makeatother

\input{fig.tex}

\preprint{{\raggedleft%
  NIKHEF 2019-030 \\
  ZU-TH 35/19 \\
  IPPP/19/61 \\
  CERN-TH-2019-113 \\
}}

\title{Associated production of a Higgs boson decaying into bottom quarks and a weak vector boson decaying leptonically at NNLO in QCD}

\author[a]{R. Gauld,}
\author[b,c]{A. Gehrmann--De Ridder,}
\author[d]{E.~W.~N.~Glover,}
\author[e]{A. Huss,}
\author[b]{I. Majer.}

\affiliation[a]{Nikhef Theory Group, Science Park 105, 1098 XG Amsterdam, The Netherlands}
\affiliation[b]{Institute for Theoretical Physics, ETH, CH-8093 Z\"urich, Switzerland}
\affiliation[c]{Department of Physics, University of Z\"urich, CH-8057 Z\"urich, Switzerland}
\affiliation[d]{Institute for Particle Physics Phenomenology, Durham University,  Durham DH1 3LE, UK}
\affiliation[e]{Theoretical Physics Department, CERN, CH-1211 Geneva 23, Switzerland}

\emailAdd{rgauld@nikhef.nl}
\emailAdd{gehra@phys.ethz.ch}
\emailAdd{e.w.n.glover@durham.ac.uk}
\emailAdd{alexander.huss@cern.ch}
\emailAdd{majeri@phys.ethz.ch}

\abstract{%
We present the calculation of next-to-next-to-leading order (NNLO) corrections in perturbative QCD for the production of a Higgs boson decaying into a pair of bottom quarks in association with a leptonically decaying weak vector boson: $\Pp\Pp \to V \PH + X \to \Pl\Pal\;\Pqb\Paqb + X$.
We consider the corrections to both the production and decay sub-processes, retaining a fully differential description of the final state including off-shell propagators of the Higgs and vector boson.
The calculation is carried out using the antenna subtraction formalism and is implemented in the \NNLOJET framework.
Clustering and identification of \Pqb-jets is performed with the flavour-$k_t$ algorithm and results for fiducial cross sections and distributions are presented for the LHC at $\sqrt{s}=\SI{13}{\TeV}$.
We assess the residual theory uncertainty by varying the production and decay scales independently and provide scale uncertainty bands in our results, yielding percent-level accurate predictions for observables in this Higgs production mode computed at NNLO.
Confronting a naïve perturbative expansion of the cross section against the customary re-scaling procedure to a fixed branching ratio reveals that starting from NNLO, the latter could be inadequate in estimating missing higher-order effects through scale variations.
}

\begin{document}
\maketitle
\clearpage
\flushbottom

\input{VH-sec1}

\input{VH-sec2}

\input{VH-sec3}

\input{VH-sec4}

\input{VH-sec5}

\acknowledgments
We are grateful to Gavin Salam for useful discussions clarifying the implementation of the flavour-$k_t$ algorithm and to Raoul Röntsch for providing numbers for a comparison against ref.~\cite{Caola:2017xuq}.
The authors also thank Xuan Chen, Juan Cruz-Martinez, James Currie, Thomas Gehrmann, Marius Höfer, Tom Morgan, Jan Niehues, Joao Pires, Duncan Walker, and James Whitehead for useful discussions and their many contributions to the \NNLOJET code.
This research was supported in part by the UK Science and Technology Facilities Council under contract ST/G000905/1, by the Swiss National Science Foundation (SNF) under contract 200021-172478, and by the Dutch Organisation for Scientific Research (NWO) through the VENI grant 680-47-461.

\clearpage
\appendix

\input{VH-AppendixA}

\input{VH-AppendixB}

\bibliography{VH}

\end{document}

%% file: VH-sec1.tex
\section{Introduction}
\label{sec:intro}

One of the highest priorities of the LHC physics programme is the detailed exploration of the mechanism of electroweak symmetry breaking that predicts the existence of the Higgs boson and its interactions with the fermions and gauge bosons of the Standard Model (SM).
In July 2012, the ATLAS and CMS Collaborations at the LHC reported the discovery of a resonance with a mass close to \SI{125}{\GeV}~\cite{Aad:2c012tfa,Chatrchyan:2012xdj}.
At the current level of accuracy, the discovered particle proves to be consistent with the Higgs boson predicted by the SM but the limited precision of some of the measurements still leaves room for possible alternative interpretations beyond the SM.
Measurements of various properties of the Higgs boson have been carried out since then.
One of the main goals of the completed Run~II  at $\sqrt{s} = \SI{13}{\TeV}$ and the future Run~III at $\sqrt{s} = \SI{14}{\TeV}$ of the LHC is to test the coupling strength of the discovered Higgs-like particle to known SM particles through the study of a variety of processes at these increased luminosity and collisions energies.

The production of a Higgs boson ($\PH$) in association with either a $\PWpm$ or a $\PZ$ boson and possible hadronic jets --- also known as \emph{Higgs Strahlung} --- is among the most promising class of channels that can lead to the accurate determination of the Higgs-boson couplings.
These were also the channels that were mainly probed during the search for a light Higgs boson at the Tevatron; and the observation of excess events at the Tevatron turned out to be consistent with the observed Higgs boson at the LHC~\cite{Aaltonen:2012qt}.

At LHC energies the $V\PH$ processes are the third ($V=\PWpm$) and fourth ($V=\PZ$) largest production channels after the dominant gluon--gluon and vector-boson-fusion ones.
These classes of Higgs production modes provide the opportunity to probe the gauge-boson--Higgs vertex ($VV\PH$) separately for $V=\PWpm$ and $V=\PZ$.
Moreover, a second and particularly relevant feature of the $\Pp\Pp \to V\PH$ process is the possibility to study the decay of a Higgs boson into a pair of bottom--antibottom quarks.
This decay is extremely important to measure since it provides a direct measurement of the Higgs coupling to fermions, thereby testing the mechanism of fermion mass generation in the SM.
Furthermore, since this decay mode dominates the total width of the Higgs boson, the uncertainty on this branching ratio enters into other studies as well, for instance in measurements of the decay of the Higgs boson to invisible final states, which are relevant for dark matter
searches~\cite{Aad:2014iia}.
Such a decay is hard to measure in inclusive Higgs production through the leading production modes like the gluon--gluon or vector-boson-fusion channels due to the presence of enormous QCD backgrounds.
In the Higgs Strahlung process the presence of a vector boson decaying leptonically provides a clean experimental signature and means experimental analyses related to $V\PH$ production have a manageable background.

Direct searches for the SM Higgs boson through $V\PH$ production and $\PH\to\Pqb\Paqb$ decay has been carried out at the LHC at centre-of-mass energies of \SI{7}{\TeV}, \SI{8}{\TeV}, and \SI{13}{\TeV}.
While the use of Run~I data at $\sqrt{s} =$ \SIlist{7;8}{\TeV} by the ATLAS and CMS Collaborations was not able to firmly establish the discovery of the SM-like Higgs boson through this channel~\cite{Aad:2012gxa,Chatrchyan:2013zna}, the use of Run~II data at $\sqrt{s} = \SI{13}{\TeV}$ enabled to do so.
In 2017, The LHC experiments~\cite{Aaboud:2018zhk,Sirunyan:2018kst} announced the observation of a SM Higgs-like particle decaying to a pair of bottom--antibottom quarks precisely through this Higgs Strahlung production channel with a significance of \numlist{5.6;5.3} standard deviations for CMS and ATLAS respectively.

In view of prospective measurements of Higgs Strahlung final states including data from Run~II and~III at the LHC, it is of crucial importance to have precise theoretical predictions for cross sections and differential distributions in the kinematic regions probed by the experiments.
This includes in particular QCD effects in both the production and in the decay of the Higgs boson into a bottom-quark pair.

The present status of theoretical predictions for observables related to $V\PH$ production with a vector boson decaying leptonically and a Higgs boson decaying into a bottom--antibottom quark pair can be summarised as follows:

The total inclusive cross section for associated $V\PH$ production is known at NNLO QCD precision.
It is available through the numerical program VH@NNLO~\cite{Brein:2012ne} whose ingredients have been reported in refs.~\cite{Harlander:2002wh,Brein:2011vx}.
The electroweak corrections to the total cross section are known at NLO~\cite{Ciccolini:2003jy,Denner:2011id}.
Differential distributions have also been computed at NNLO QCD, including the computation of $\PH\to\Pqb\Paqb$ decay at different orders.
In refs.~\cite{Ferrera:2014lca,Ferrera:2011bk,Campbell:2016jau}, the Higgs decay has been included at NLO while it is included up to NNLO in refs.~\cite{Ferrera:2017zex,Caola:2017xuq}.
In addition, the fully differential decay rate for $\PH\to\Pqb\Paqb$ known so far at NNLO QCD~\cite{Anastasiou:2011qx,DelDuca:2015zqa} has recently been computed at N$^3$LO accuracy in ref.~\cite{Mondini:2019gid}, although jet-flavour is not identified in this calculation.
The combination of fixed-order QCD computations with parton showers have also been the subject of phenomenological studies~\cite{Hamilton:2012rf,Luisoni:2013kna,Astill:2018ivh}.

Fully differential NNLO predictions for $V\PH$ observables obtained via the combination of Higgs production and decay to bottom--antibottom processes have been presented in ref.~\cite{Ferrera:2017zex} (for $V=\PZ, \PWp$) and in ref.~\cite{Caola:2017xuq} (for $V=\PWm$).
These computations have essential features in common: at parton level, both consider massless $\Pqb$-quarks except in the Higgs Yukawa coupling and use the same flavour-$k_t$ algorithm~\cite{Banfi:2006hf} to define $\Pqb$-jets.
Furthermore, the Higgs decay is treated in the narrow-width approximation and the Higgs Yukawa coupling $y_\Pqb$ is computed at fixed scale $\mu=m_{\PH}$.
Scale variations are only considered in the production sub-process using the central scale choice $\mu=M_{V\PH}$.

The aforementioned computations differ instead in the theoretical framework employed to regulate infrared divergences at NNLO level: in ref.~\cite{Ferrera:2017zex} the $q_T$-subtraction formalism~\cite{Catani:2007vq} is used for the $V\PH$ production cross section combined with the CoLoRFulNNLO subtraction method~\cite{DelDuca:2016ily} for the $\PH\to\Pqb\Paqb$ decay.
In ref.~\cite{Caola:2017xuq} the nested soft-collinear subtraction scheme~\cite{Caola:2017dug,Caola:2019pfz} is used (an extension of the residue subtraction scheme~\cite{Czakon:2014oma}) in both production and decay sub-processes.

It is the aim of this paper to present a computation of $V\PH$ observables for all three processes $(V=\PZ,\PWpm)$ including NNLO corrections to both production and decay sub-processes.
Our goal  is to yield a fully differential description of the final states, i.e.\ including the decays of the vector boson into leptons and the Higgs boson into bottom quarks with off-shell propagators of the vector- and Higgs-boson.
The NNLO corrections to both production and decay sub-processes are calculated using the antenna subtraction formalism~\cite{GehrmannDeRidder:2005cm,GehrmannDeRidder:2005aw,GehrmannDeRidder:2005hi,Daleo:2006xa,Daleo:2009yj,Boughezal:2010mc,Gehrmann:2011wi,GehrmannDeRidder:2012ja,Currie:2013vh} implemented within the \NNLOJET framework~\cite{Currie:2018oxh}.

The structure of this paper is as follows: in section~\ref{sec:flavour}, we provide an overview explaining how flavour-dependent observables are computed at fixed-order accuracy within the parton level event generator \NNLOJET.
A detailed description of the jet-algorithm used to achieve this goal, as well as its application to the $V\PH$ process are also specified.
In section~\ref{sec:details}, we present the details of the $V\PH$ calculation giving explicitly the different ingredients appearing in production and decay sub-processes up to NNLO level.
Section~\ref{sec:results} contains our results for the fiducial cross sections and differential observables related to $V\PH$ production in $\Pp\Pp$ collisions at $\sqrt{s}=\SI{13}{\TeV}$.
Those include, for the first time, scale uncertainty estimations related to the separate variation of production and decay scales at each order in $\alphas$.
Two appendices are enclosed: in appendix~\ref{sec:eventag} the impact of different criteria for defining the net flavour of jets is studied for a number of relevant NNLO distributions in $V\PH$ production.
Appendix~\ref{sec:fixbr} is dedicated to a comparison of results obtained using two different expressions of the cross section, including either a fixed branching ratio $\text{Br}(\mathrm{H\to \Pqb \bar{b}})$ as used previously in refs.~\cite{Caola:2017xuq,Ferrera:2017zex,Campbell:2016jau}, or not, as in this paper.

%% file: VH-sec2.tex
\section{Flavour tagging of jets}
\label{sec:flavour}

The goal of this work is to provide fixed-order predictions for the hadron-level process $\Pp\Pp\to\Pl\Pal\,+2\,\Pqb\text{-jets}+X$, i.e.\ yielding a final state which contains flavour-tagged bottom-quark jets (\Pqb-jets) and (charged) leptons.
The presence of two identified \Pqb-jets with a combined invariant mass consistent with $m_\PH$ allows us to associate this final state with the underlying process $\Pp\Pp \to V\PH \,+\, X \to \Pl\Pal\;\Pqb\Paqb \,+\, X$.
The identification of jet flavour is an essential component of any experimental analysis of this process, which is required to reduce otherwise overwhelming background processes.
It is therefore also an integral part of the requirements needed to obtain the corresponding theoretical predictions.

The computation of such observables at fixed order requires the application of a flavour-sensitive jet algorithm that --- besides reconstructing flavour-insensitive properties such as four-momenta --- identifies the flavour of the reconstructed jets based on some well-defined (infrared-safe) criteria~\cite{Banfi:2006hf}.
The application of such an algorithm requires a tracking of the flavour of individual partons, which appear in the partonic cross section at each perturbative order.

In the following, we provide a description of how this is achieved within the parton-level event generator \NNLOJET.
The discussion is however not specific to the use of the antenna subtraction formalism to regulate infrared divergences occurring in partonic sub-processes beyond LO.
In addition, as the application of a flavour-sensitive jet algorithm is not standard (although required from the point of view of massless fixed-order computations) for either theory or experimental communities, we also give a brief overview of the algorithm used for these computations.
This section is concluded by providing specific details of the jet clustering implementation relevant for the results presented here regarding the computation of NNLO observables for $V\PH$ production.

\subsection{Flavour dressing}

The first step towards computing flavour-dependent jet observables is to ensure that the jet algorithm has access to both momentum and flavour information when evaluating the contributions from matrix elements and subtraction terms.
To address this issue within \NNLOJET, an additional ``flavour-dressing" layer that tracks the flavours of all amplitudes as well as reduced matrix elements appearing in subtraction terms has been implemented.

To illustrate how this proceeds, we consider the construction of a generic NLO-type cross section for an $n$-body final state initiated by the two partons $i$ and $j$. Following the notation of ref.~\cite{Currie:2013vh}, we may write the contribution to the partonic cross section as
\begin{align} \label{eq:NLO}
	\rd\hat{\sigma}_{ij,\NLO} & =
	\int_{n+1} \left[ \rd\hat{\sigma}^{R}_{ij,\NLO} - \rd\hat{\sigma}^{S}_{ij,\NLO}\right]
	+ \int_{n} \left[ \rd\hat{\sigma}^{V}_{ij,\NLO} - \rd\hat{\sigma}^{T}_{ij,\NLO}\right] ,
\end{align}
where the superscripts $R$, $S$, $V$, $T$ indicate the real, real-subtraction, virtual, and virtual-subtraction terms, respectively.

As an example of the flavour-dressing procedure for the amplitudes, we consider the real-emission cross section (omitting the sum over potential colour orderings) which takes the general form
\begin{align}
	\rd \hat{\sigma}^R_{ij,\NLO} & =
	\mathcal{N}^R_{\NLO} \;\rd \Phi_{n+1} \left(\left\{p_3,\ldots,p_{n+3}\right\}; p_{1}, p_{2}\right) \; \frac{1}{S_{n+1}} \nonumber \\
	                             & \times \left[
	M^0_{n+3}   \left(\left\{p_{n+3}\right\}, \left\{f_{n+3}\right\} \right) \;
	J_n^{(n+1)}\! \left(\left\{p_{n+1}\right\}, \left\{f_{n+1}\right\} \right) \right] .
\end{align}
We denote the final-state symmetry factor by $S_{n+1}$, the normalisation factor (which includes constants, couplings, colour factors) by $\mathcal{N}^R_{\NLO}$, the $2\to n+1$ particle phase space by $\rd \Phi_{n+1}$, and the momentum of the partons $i,j$ by $p_{1,2}$.
The partial squared amplitude $M^0_{n+3}$ is evaluated with the momentum set $\left\{p_{n+3}\right\}$ and a corresponding flavour set $ \left\{f_{n+3}\right\}$.
The flavour-sensitive jet algorithm $J_n^{(n+1)}$ builds $n$ jets from $n+1$ final-state partons which carry momentum and flavour labelled by the sets $\left\{p_{n+1}\right\}$ and $\left\{f_{n+1}\right\}$ respectively.

The real subtraction cross section can be written in a similar fashion:
\begin{align}
	\rd \hat{\sigma}^S_{ij,\NLO} & =
	\mathcal{N}^R_{\NLO} \,\sum_{k} \rd \Phi_{n+1} \left(\left\{p_3,\ldots,p_{n+3}\right\}; p_{1}, p_{2}\right) \; \frac{1}{S_{n+1}} \nonumber \\
	                             & \times \left[
	X_3^0(\cdot,k,\cdot) \;
	M^0_{n+2} \left(\{\tilde{p}_{n+2}\}, \{\tilde{f}_{n+2}\} \right) \;
	J_n^{(n)}\! \left(\{\tilde{p}_n\}, \{\tilde{f}_n\}         \right)  \right] ,
\end{align}
where the index $k$ runs over all possible unresolved partons in $\rd \hat{\sigma}^R_{ij,\NLO}$ and $X_3^0(\cdot,k,\cdot)$ denotes the three-parton antenna function that factorises from the associated reduced squared matrix-element $M^0_{n+2}$.
In this case, the jet algorithm acts upon mapped final-state momentum and flavour sets $\{\tilde{p}_n\}$ and $\{\tilde{f}_n\}$ associated with the reduced squared matrix element $M^0_{n+2}$.
As the total subtraction cross section must take into account all possible unresolved limits of parton $k$, this cross section may be composed of multiple flavour structures.
The subtraction method is only effective if the evaluation of flavour-dependent observables in both the real and real-subtraction cross sections match in all possible unresolved limits. This is only ensured if an infrared-safe flavour-sensitive jet algorithm is applied.

To construct the NLO cross section according to eq.~\eqref{eq:NLO}, a similar procedure must also be applied to both virtual and virtual-subtraction (in the antenna formalism, these include integrated subtraction and mass-factorisation contributions) cross sections.
This construction is obtained in a similar fashion, by tracking both the momentum and flavour sets associated to all partial squared amplitudes and reduced squared matrix elements appearing in these contributions and then applying the flavour-sensitive jet algorithm to the subset of final-state particles within these sets.
To allow the computation of flavour-dependent jet observables at NNLO, the same ideas extend to one order higher and this flavour-dressing procedure is applied to all NNLO-type parton level contributions and their corresponding subtraction terms.

\subsection{Flavoured-jet algorithm}
\label{sec:jet}

Throughout this work jets are reconstructed with the flavour-$k_t$ algorithm, which provides an infrared-safe definition of jet flavour.
The main difference with respect to a native jet algorithm is that the clustering of particles relies on both momentum and flavour information of the input pseudo-jets.
For completeness, we summarise the main steps of the algorithm for hadron--hadron collisions originally presented in ref.~\cite{Banfi:2006hf} (also summarised in ref.~\cite{Banfi:2007gu}).

The algorithm proceeds by assigning a net flavour to all pseudo-jets or jets based on their quark flavour content, attributing $+1$ ($-1$) if a quark (antiquark) of the flavour under consideration is present.
In an experimental context, the presence of a quark flavour could be inferred from a fully/partially reconstructed hadron.
A criterion is then applied to these objects to determine if they carry flavour, possible examples being: the net flavour (sum of quarks and antiquarks); or the net flavour modulo two.
Objects are considered to carry flavour if they carry non-zero values of this criterion.
The algorithm then proceeds by constructing distance measures for pairs of all final-state pseudo-jets $i$ and $j$ ($d_{ij}$) as well as beam distances ($d_{iB}$ and $d_{i\bar{B}}$).
These (flavour-dependent) distances are defined as
\begin{equation}
	\label{eq:flavourkt_meas}
	d_{ij} = \frac{\Delta y_{ij}^2 + \Delta \phi_{ij}^2}{R^2}
	\begin{cases}
		\max(k_{ti}, k_{tj})^{\alpha}\,\min(k_{ti}, k_{tj})^{2-\alpha} & \text{softer of } i, j \text{ is flavoured,}   \\
		\min(k_{ti}, k_{tj})^{\alpha}                                  & \text{softer of } i, j \text{ is unflavoured,} \\
	\end{cases}
\end{equation}
and
\begin{equation}
	\label{eq:flavourkt_beam}
	d_{i\parenbar{B}} =
	\begin{cases}
		\max(k_{ti}, k_{t\parenbar{B}}(y_i))^{\alpha}\,\min(k_{ti}, k_{t\parenbar{B}}(y_i))^{2-\alpha} & \text{softer of } i, j \text{ is flavoured,}   \\
		\min(k_{ti}, k_{t\parenbar{B}}(y_i))^{\alpha}                                                  & \text{softer of } i, j \text{ is unflavoured.} \\
	\end{cases}
\end{equation}
In these definitions, $k_{ti}$ and $k_{tj}$ are the transverse momentum of the pseudo-jets $i$ and $j$, and the rapidity difference and azimuthal angular separation between these pseudo-jets is given by $\Delta y_{ij}$ and $\Delta \phi_{ij}$, respectively.
The parameters $R$ and $\alpha$ define a class of measures for the algorithm.
The (rapidity-dependent) transverse momentum of the beam $B$ at positive rapidity $k_{tB}$, and beam $\bar{B}$ at negative rapidity $k_{t\bar{B}}$, are defined as:
\begin{align}
	k_{tB}(y)       & = \sum_i k_{ti} \left( \Theta(y_i - y) +  \Theta(y - y_i) \; \re^{y_i-y} \right), \label{eq:ktb}    \\
	k_{t\bar{B}}(y) & = \sum_i k_{ti} \left( \Theta(y - y_i) +  \Theta(y_i-y)   \; \re^{y-y_i} \right), \label{eq:ktbbar}
\end{align}
with $ \Theta(0) = 1/2$ and the index $i$ going over all pseudo-jets.

While this flavour-aware jet algorithm is substantially more complex than the flavour-blind anti-$k_t$ algorithm~\cite{Cacciari:2008gp}, its use is unavoidable in fixed-order computations based on massless quarks.
At NLO, the flavour criterion of a pseudo-jet ensures that a collinear splitting of the form $\Pg\to\Pq\Paq$ is indistinguishable from a gluon (or flavourless) jet.
The subtraction formalism presented in eq.~\eqref{eq:NLO} would already be spoiled without this criterion.
At NNLO, the flavour-dependent distance measure in eq.~\eqref{eq:flavourkt_meas} ensures that a pair of flavoured quarks originating from a wide-angle gluon splitting is clustered into a pseudo-jet \emph{before} being combined with any other (harder) pseudo-jets.
This avoids the situation where one of these soft quarks may be clustered with a hard pseudo-jet that carries zero flavour, which would lead to a definition of jet flavour sensitive to soft physics.
These are issues which are otherwise insurmountable for fixed-order computations involving massless quarks.

The flavour-$k_t$ algorithm described above is available in the \NNLOJET framework and has been validated against an independent implementation using FastJet~\cite{Cacciari:2005hq,Cacciari:2011ma}.

\subsection{Jet clustering for the \texorpdfstring{$V\PH$}{VH} process}
\label{sec:jet_vh}

The discussion of flavour dressing and the jet algorithm presented in this section are quite general and are applicable to all processes implemented within \NNLOJET.
Here we discuss a few specific points related to the application of the jet algorithm to the $V\PH$ process, which will be relevant to the results presented in later sections of this paper.

\begin{figure}
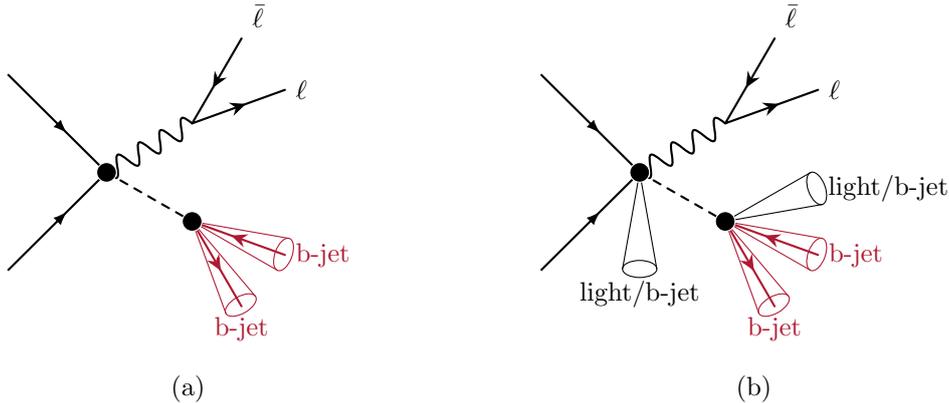

	\centering
	\begin{subfigure}[t]{0.45\textwidth}
		\centering
		\VHjets{0}{1.3cm}{0}
		\caption{}
		\label{fig:vh_jets_a}
	\end{subfigure}
	\hspace{1em}
	\begin{subfigure}[t]{0.45\textwidth}
		\centering
		\VHjets{1}{1.3cm}{0}
		\caption{}
		\label{fig:vh_jets_b}
	\end{subfigure}
	\caption{Possible event configurations characterised by the presence of two hard \Pqb-jets at LO in~(\subref{fig:vh_jets_a}) and at NNLO in~(\subref{fig:vh_jets_b}) where multiple \Pqb-jets and light jets can be emitted from the production or the decay sides.}
	\label{fig:vh_jets}
\end{figure}

The first point is that we wish to apply the flavour-$k_t$ algorithm to the partonic process $\Pq\Paq\to V\PH\to \Pl\Pal\;\Pqb\Paqb$, including NNLO QCD corrections which will be discussed in section~\ref{sec:details}.
When higher-order corrections are considered, additional light or \Pqb-quark partons can be emitted from both production and the decay sides, as illustrated in figure~\ref{fig:vh_jets_b}.
The jet clustering is performed by considering \Pqb-quarks to be flavoured (all other partons carrying zero flavour) and fully accounting for emissions from both production and decay during the jet clustering process.
While our calculation focusses on the decay sub-process $\PH\to\Pqb\Paqb$, it has been implemented in such a way that predictions for the hadronic process $\Pp\Pp\to\Pl\Pal\,+2\,\Pqc\text{-jets}+X$ can also be produced.
This may be interesting in view of possible future measurements by the LHCb Collaboration~\cite{LHCb-CONF-2016-006}.

The second point is that the definition of the transverse momentum of the beam is altered to account for the presence of a leptonically decaying gauge-boson.
This is done by modifying eq.~\eqref{eq:ktb} according to
\begin{align}
	\tilde{k}_{tB}(y) & = k_{tB}(y) + E_{t,V} \left( \Theta(y_V - y) +  \Theta(y - y_V)  e^{y_V-y} \right),
\end{align}
where $E_{t,V}$ and $y_V$ are the transverse energy and rapidity of the reconstructed gauge-boson.
A similar modification to the beam transverse momentum at negative rapidity~\eqref{eq:ktbbar} is assumed.
This modification is introduced to provide a better estimate of the hardness of the beam, which can affect the clustering outcome.
One could alternatively modify the beam hardness by including the charged leptons, which may be necessary in experimental situations where the gauge-boson cannot be fully reconstructed.

The final point is related to our choice of flavour criterion during the clustering process.
We have chosen to define the flavour of pseudo-jets to be the net-flavour of its constituents modulo two, which means that all pseudo-jets which contain an even flavour content are considered to have zero net-flavour.
The motivation for this choice is that, in our opinion, it is the most feasible realisation of the flavour-$k_t$ algorithm experimentally.
Focussing on the case of \Pqb-jets, the main consideration is that most experimental approaches to flavour tagging are sensitive only to the absolute flavour~\cite{Aaij:2015yqa,Aad:2015ydr,Sirunyan:2017ezt} (and do not additionally charge tag the jets).
All implementations of the algorithm must consider the combination of a $\Pqb\Paqb$-quark pair (or equivalently a $\PB\PBb$-hadron pair) as carrying zero flavour, as required to guarantee its infrared safety as discussed above.
Therefore, in the absence of charge tagging, any (pseudo)-jet which contains the presence of an even number of \Pqb (\PB) and/or \Paqb (\PBb) quarks (hadrons) should also be considered to carry zero flavour, as experimentally these signatures are indistinguishable.

The charge-tagging of flavoured jets is also possible~\cite{Aaij:2014ywa}, for example in the presence of semi-leptonic \PB-hadron decays.
However, the drawback is a large reduction in event statistics (roughly an order of magnitude for each \Pqb-jet, as the branching fraction ${\mathrm{Br}(\PB\to \Pl+X)} \approx 10\%$) with little informational gain.
Accordingly, to present our results for NNLO observables related to $V\PH$ production in section~\ref{sec:results}, we shall use the version of the flavour-$k_t$ algorithm where all even-tagged (pseudo)-jets carry zero flavour. We further provide an examination of the impact of the even-tag exclusion in the shape and normalisation of flavour sensitive observables in appendix~\ref{sec:eventag}.

%% file: VH-sec3.tex
\section{Details of the calculation}
\label{sec:details}

In this section we present the main ingredients that enter the calculation of the Higgs Strahlung process at NNLO.
We establish how those building blocks are assembled to express the cross section in a factorised form in terms of production and decay sub-processes.

\subsection{General framework}
\label{sec:genframe}

We consider the process $\Pp\Pp\to V \PH \,+\,X \to \Pl\Pal\;\Pqb\Paqb \,+\,X$ where the vector boson ($V$) decays leptonically and the Higgs boson (\PH) decays into a pair of bottom quarks $\Pqb\Paqb$.
We compute NNLO QCD observables related to these reactions by including corrections up to order $\alphas^2$ in both production and decay sub-processes.
This enables us to express the fully differential cross section at the $k$th order in a factorised form given as
\begin{align}
	\label{eq:VH1}
	\rd\sigma^{\N{k}\LO} & =
	\sum_{\substack{i,j=0    \\ i + j \le k}}^{k}
	\rd \sigma^{(i)}_{V\PH} \times \rd \sigma^{(j)}_{\PH\to\Pqb\Paqb} \,.
\end{align}
The term $\rd\sigma^{(i)}_{V\PH}$, which corresponds to the production part, comprises the vector propagator and the leptonic decay $V\to\Pl\Pal$, including spin correlations between the initial-state partons and the final-state leptons.
The term denoted by $\rd\sigma^{(j)}_{\PH\to\Pqb\Paqb}$ corresponds to the decay part and includes the Higgs propagator and its subsequent decay to a bottom--antibottom quark pair.
We treat all light quarks as massless including the bottom quark with the exception of the Yukawa coupling mediating the $\PH\to\Pqb\Paqb$ decay.
In the decay the bottom quark Yukawa coupling to the Higgs boson is renormalised in the \MSbar scheme at the scale $\mu^{\text{dec.}}$, taken to be proportional to the Higgs-boson mass $m_\PH$.%
\footnote{
	It is known from the computation of the inclusive cross section that this choice of regularisation scheme leads to a reduction of the size of the QCD corrections.
}
Note that the factorised form of the associated Higgs production cross section~\eqref{eq:VH1} does not include interferences between production and decay sub-processes.
This is a valid approximation owing to the smallness of the Higgs decay width, which further formed the basis of the narrow-width approximation (NWA) as used in previous calculations.

Up to $\cO(\alphas^2)$, the cross section may then be written as
\begin{align}
	\rd \sigma^{\NNLO}
	 & = \rd \sigma_{V\PH}^{(0)} \times \left( \rd \sigma^{(0)}_{\PH\to\Pqb\Paqb} + \rd \sigma^{(1)}_{\PH\to\Pqb\Paqb}  + \rd \sigma^{(2)}_{\PH\to\Pqb\Paqb} \right) \nonumber \\
	 & + \rd \sigma_{V\PH}^{(1)} \times \left( \rd \sigma^{(0)}_{\PH\to\Pqb\Paqb} + \rd \sigma^{(1)}_{\PH\to\Pqb\Paqb} \right) \nonumber                                       \\
	 & + \rd \sigma_{V\PH}^{(2)} \times \left( \rd \sigma^{(0)}_{\PH\to\Pqb\Paqb} \right) \, .
	\label{eq:VH2}
\end{align}
We note that this formulation of the NNLO cross section does not contain the Higgs-boson branching ratio into \Pqb quarks given as $\text{Br}(\PH\to\Pqb\Paqb) = \Gamma_{\PH\to\Pqb\Paqb}/\Gamma_{\PH}$.
This is in contrast to previous calculations for the $V\PH$ process at NNLO, either considering the decay sub-process at NLO~\cite{Ferrera:2014lca,Ferrera:2011bk,Campbell:2016jau} or NNLO~\cite{Ferrera:2017zex,Caola:2017xuq}, which all employed a scaled variant of the cross section incorporating the Higgs-boson branching ratio at a fixed value and thus not subject to an $\alphas$ expansion.
It is worth mentioning that this scaled variant of the cross section was essential in describing the data using fixed-order predictions at LO and NLO~\cite{Ferrera:2011bk,Ferrera:2014lca}.
With this formulation, the LO predictions have the correct normalisation; NLO corrections are kept small and have a small residual theoretical uncertainty.
If computed up to order $\alphas^2$, we here argue that the need of such scaling factors in the formulation of the cross section becomes questionable.

In appendix~\ref{sec:fixbr}, we will further elaborate on this matter and compare the results obtained with both approaches for the fiducial cross sections up to order $\alphas^2$.
We find that a consistent treatment of theoretical uncertainties outweighs the precision gain that one might obtain by scaling the cross section to a fixed branching ratio, if the cross section is computed including NNLO corrections in both production and decay parts.
This further motivates the simpler formulation of the cross section given above in eq.~\eqref{eq:VH2} where no scaling factors are applied.
This will be our default setup throughout this work and for the results presented in section~\ref{sec:results}.

As a validation of our calculation, we performed a comparison to the results of ref.~\cite{Caola:2017xuq} by adopting their calculational setup and found perfect agreement with the reported values for the total cross sections in Table~I at each perturbative order in $\alphas$.

\subsection{Production and decay parts up to \texorpdfstring{$\cO(\alphas^2)$}{O(alphas**2)} }

Based on our master formula~\eqref{eq:VH2} for the $V\PH$ process at NNLO, we specify the individual ingredients of the production and decay parts in the following and describe how they are assembled to the final prediction for the Higgs Strahlung process.

\subsubsection{Production parts}

Up to order $\alphas$, only one type of contribution enters the associated Higgs production cross section, which is given by Drell--Yan-like diagrams with a subsequent Higgs emission from the gauge-boson leg.
Starting from $\cO(\alphas^2)$, additional quark-loop induced contributions arise.
These can be treated independently from the aforementioned Drell--Yan-type ones as the relevant Feynman amplitudes are individually gauge invariant.
In the following, we describe these two production modes one after the other.

\paragraph{Drell--Yan-type:}
These contributions arise from the Drell--Yan-like production of a virtual \PWpm or \PZ boson, which then splits into a real vector boson and a Higgs particle.
In our calculation we include them up to $\cO(\alphas^2)$ using off-shell amplitudes that effectively treat both the directly produced vector boson and the vector boson that decays leptonically as virtual particles.
\begin{figure}
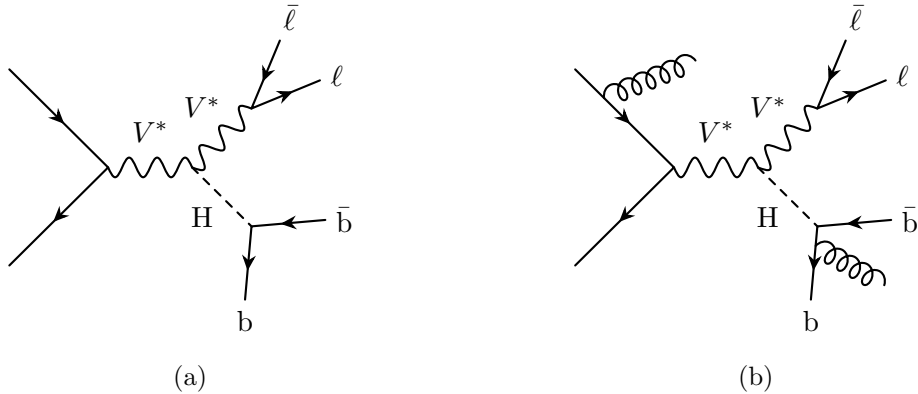

	\centering
	\begin{subfigure}[t]{0.45\textwidth}
		\centering
		\DrellYan{0}{0}{1.3cm}{0}
		\caption{}
		\label{fig:drell-yan_a}
	\end{subfigure}
	\hspace{1em}
	\begin{subfigure}[t]{0.45\textwidth}
		\centering
		\DrellYan{1}{1}{1.3cm}{0}
		\caption{}
		\label{fig:drell-yan_b}
	\end{subfigure}
	\caption{Examples of Feynman diagrams entering the Drell--Yan type contributions at~(\subref{fig:drell-yan_a}) LO and at~(\subref{fig:drell-yan_b}) NNLO.
    Production and decay parts have an additional final state gluon in~(\subref{fig:drell-yan_b}) compared to case~(\subref{fig:drell-yan_a}).
		Both vector bosons and the Higgs boson are treated in an off-shell manner, as explained in the main text.}
	\label{fig:drell-yan}
\end{figure}
Representative Feynman diagrams for this production mode are illustrated in figure~\ref{fig:drell-yan} at LO~(\subref{fig:drell-yan_a}) and NNLO~(\subref{fig:drell-yan_b}).

These contributions only involve the square of Drell--Yan-like amplitudes and the infrared singularities are dealt with using the NNLO antenna subtraction formalism~\cite{GehrmannDeRidder:2005cm}.
The subtraction terms can be readily constructed based on the NNLO calculation for the Drell--Yan processes, which are available within the \NNLOJET framework.

\paragraph{Top-quark-loop induced:}
Starting from $\cO(\alphas^2)$, new types of diagrams induced by quark loops must be taken into account for the $V\PH$ production process.
Depending on whether the gauge boson and/or the Higgs boson couple to the quark loop, these contributions can be classified into three categories:
\begin{enumerate}
	\myitem{(a)}\label{item:ri} A class of amplitudes with no vector boson coupling to the quark loop.
	        As such, this class contributes to all Higgs Strahlung processes involving either \PZ or \PWpm bosons.
	\myitem{(b)}\label{item:self-zh-i} A second class of amplitudes that is only present for $\PZ\PH$ production where the gauge boson as well as the Higgs boson directly couple to the quark loop.
	\myitem{(c)}\label{item:self-zh-ii} Finally, the class of amplitudes where only the \PZ boson attaches to the quark loop while the Higgs boson is emitted from the external massive gauge-boson leg.
\end{enumerate}
\begin{figure}
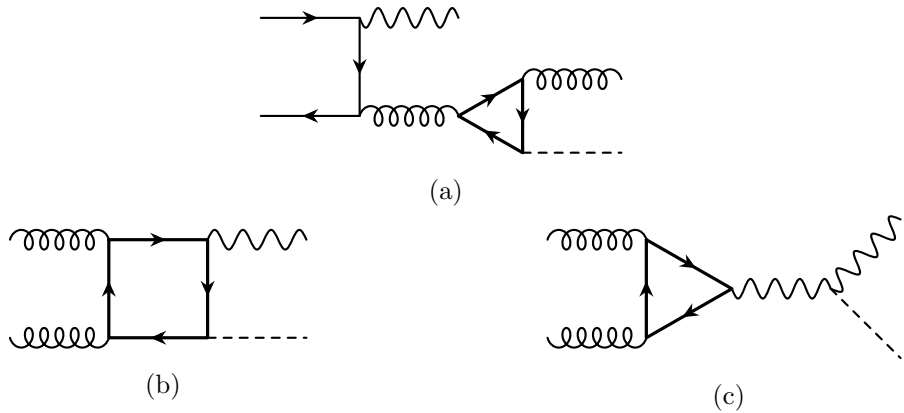

	\centering
	\begin{subfigure}[t]{0.45\textwidth}
		\centering
		\TopRI{1.3cm}{0}
		\caption{}
		\label{fig:ri}
	\end{subfigure}

	\begin{subfigure}[t]{0.45\textwidth}
		\centering
		\TopSelfZHI{1.3cm}{0}
		\caption{}
		\label{fig:top-self-zh-i}
	\end{subfigure}
	\hspace{1em}
	\begin{subfigure}[t]{0.45\textwidth}
		\centering
		\TopSelfZHII{1.3cm}{0}
		\caption{}
		\label{fig:top-self-zh-ii}
	\end{subfigure}
	\caption{Representative Feynman diagrams of the heavy-quark-loop-induced amplitudes at $\cO(\alphas^2)$ included in our calculation.
		Figure~(\subref{fig:ri}) depicts an $R_I$-type amplitude, which is present for both $\PZ\PH$ and $\PW\PH$ production channels.
		Figures~(\subref{fig:top-self-zh-i},\subref{fig:top-self-zh-ii}) illustrate representative gluon--gluon induced heavy quark loop amplitudes, which are only present for $\PZ\PH$ production.}
	\label{fig:top-loop-inc}
\end{figure}
Representative Feynman diagrams for each part are shown in figures~\ref{fig:top-loop-inc}(\subref{fig:ri}--\subref{fig:top-self-zh-ii}), respectively, where we have omitted the Higgs decay for clarity.
The contribution to the cross section either arises through the square of these diagrams (e.g.\ for the gluon--gluon-induced channels) or though the interference with Drell--Yan-type amplitudes.
Note that all quark-loop-induced contributions are both infrared and ultraviolet finite and thus no subtraction procedure is needed in their evaluation.

For cases~\ref{item:ri} and \ref{item:self-zh-i} where the Higgs boson directly couples to the quark loop, we only consider the top quark running inside the loop.
They constitute the dominant contribution in this class and are proportional to the second power of the top Yukawa coupling $y_\Pqt$.
The omission of the light-quark --- including the bottom-quark --- amplitudes is justified by their much smaller Yukawa couplings.

In case~\ref{item:self-zh-ii} on the other hand, the Higgs boson does not couple directly to the quark loop and we have to consider all quark flavours inside the loop.
For the quarks of the first two generations: $\Pq=\Pqu,\,\Pqd,\,\Pqs,\,\Pqc$, the corresponding amplitudes cancel due to the fact that an equal count of up- and down-type quark flavours are evaluated.
This cancellation is spoiled in the case of the third generation due to the non-vanishing mass of the top quark.
As a result, both the top- and bottom-loop components are included.

The complete $\cO(\alphas^2)$ top-loop-induced contributions were computed for on-shell vector bosons in ref.~\cite{Brein:2011vx}, relying in some cases on the infinite-top-mass approximation.
In our NNLO calculation we include those that are known in the exact theory and numerically sizeable but omit those which are only known in the infinite-top-mass limit.
Specifically, for the NNLO contributions associated with diagrams~\ref{item:ri}, we include diagrams with top-quark-loop insertions onto an external gluon line.
The related amplitudes are referred to as $R_I$ in ref.~\cite{Brein:2011vx} and have been included in the previous computations~\cite{Ferrera:2017zex,Caola:2017xuq}.%
\footnote{
	We did not include the two-loop amplitudes from this class as they are currently not known in the full theory but only in the infinite-top-mass limit.
	Diagrammatically, these would be given by
	\newline\bigskip
	\centerline{\TopVI{1cm}{0}}
	and are referred to as $V_I$ in ref.~\cite{Brein:2011vx}.
	The numerical impact to the total NNLO cross section is estimated to be below the percent level.
	This contribution was omitted in ref.~\cite{Ferrera:2017zex} but kept in ref.~\cite{Caola:2017xuq}.
}
Regarding the amplitudes of class~\ref{item:self-zh-i} and~\ref{item:self-zh-ii}, which are exclusively present in $\PZ\PH$ production, we only include the gluon--gluon-induced channels shown in figures~\ref{fig:top-loop-inc}(\subref{fig:top-self-zh-i},\subref{fig:top-self-zh-ii}).
Phenomenologically, they represent the dominant component among the top-loop-induced contributions due to the large gluon luminosity at the LHC and were also considered in the previous calculations at NNLO.%
\footnote{
	The contributions that we omitted from this class are are given by diagrams of the following type:
	\newline\bigskip
	\centerline{\TopRII{1cm}{0} \hspace{2em} \TopVII{1cm}{0.25cm}}
	They are denoted as $R_{II}$ and $V_{II}$ respectively in ref.~\cite{Harlander:2002wh}.
	The one-loop amplitude $R_{II}$ is known in the full theory but it merely constitutes a sub-permille effect.
	The two-loop amplitude $V_{II}$ is currently only known in the large-top-mass limit but its impact is estimated to be at the sub-percent level.
	These contributions were also omitted in ref.~\cite{Ferrera:2017zex}.
}

The heavy-quark-loop-induced contributions included in our calculation have been either independently rederived or implemented using known results, in particular those given in ref.~\cite{Campbell:2016jau}.
A validation of the implementation was performed against OpenLoops amplitudes~\cite{Cascioli:2011va} and full agreement was found in all cases.

\subsubsection{Decay parts}

For the decay sub-process $\PH\to\Pqb\Paqb$, we required the corrections up to $\cO(\alphas^2)$ as indicated in our master formula~\eqref{eq:VH2}.
The corresponding amplitudes at one- and two-loop level were obtained from the analytic expressions of refs.~\cite{Anastasiou:2011qx,DelDuca:2015zqa} and were decomposed into different colour levels according to antenna formalism conventions.
A validation of all one-loop amplitudes was performed against the OpenLoops library~\cite{Cascioli:2011va}, yielding full agreement.
In addition, subtraction terms capturing the infrared singularities are required.
Those have been constructed for the Higgs decay up to order $\cO(\alphas^2)$ for the present computation.
Checks for the correct divergent behaviour in all single- and double-unresolved limits have been performed in order to ensure the proper cancellation of singularities in the real-emission corrections as well as the cancellation of poles against the virtual amplitudes.

The decay sub-process up to NNLO only enters in eq.~\eqref{eq:VH2} when combined with the Drell--Yan-type production parts.
For the top-quark-loop induced contributions, which are already of $\cO(\alphas^2)$, the decay only needs to be considered at tree level.

%% file: VH-sec4.tex
\section{Numerical results}
\label{sec:results}

In this section we present phenomenological results obtained for the different $V\PH$ processes using our implementation in the parton-level event generator~\NNLOJET.
We first summarise the general setup in section~\ref{sec:setup} and move on to discuss the integrated fiducial cross sections obtained within this setup in section~\ref{sec:fiducial}.
We devote section~\ref{sec:scale} to validating the scale dependence of our numerical results and present differential distributions for flavour-sensitive observables in section~\ref{sec:results_dist}.

\subsection{General setup}
\label{sec:setup}

We provide predictions for proton--proton collisions at $\sqrt{s} = \SI{13}{\TeV}$ using the parton distribution function \verb|NNPDF31_nnlo_as_0118| provided via the LHAPDF library~\cite{Buckley:2014ana}.
Each event was required to contain at least two \Pqb-jets with transverse momentum $p_{\bot,\Pqb} > \SI{25}{\GeV}$ and rapidity $|y_{\Pqb}| < \num{2.5}$.
Charged leptons were required to have a transverse momentum above $p_{\bot,\Pl} > \SI{15}{\GeV}$ and for their rapidity to satisfy $|y_{\Pl}| < \num{2.5}$.
For the $\PWpm\PH$ processes, we additionally demanded a minimum missing transverse energy of $E_{\bot,\text{miss}} > \SI{15}{\GeV}$ to identify the neutrino in the final state.
We used the flavour-$k_t$ algorithm with an even-tag exclusion to cluster \Pqb-jets as described in sections~\ref{sec:jet} and \ref{sec:jet_vh} with the parameters $R=0.5$ and $\alpha=2$.

We employed the $G_\mu$-scheme for the electroweak input parameters and the full set of independent parameters entering the computation are given by
\begin{align}
	m_{\PZ}              & = \SI{91.1876}{\GeV} ,           &
	m_{\PW}              & = \SI{80.385}{\GeV} ,            &
	m_{\PH}              & = \SI{125.09}{\GeV} ,
	\nonumber                                                 \\
	\Gamma_{\PZ}         & = \SI{2.4952}{\GeV} ,            &
	\Gamma_{\PW}         & = \SI{2.085}{\GeV} ,             &
	\Gamma_{\PH}         & = \SI{4.1}{\MeV} ,
	\\
	\overline{m}_{\Pqb}
	                     & = \SI{4.18}{\GeV} ,              &
	m_{\Pqt}^\text{pole} & = \SI{173.21}{\GeV} ,            &
	G_\text{F}           & = \SI{1.1663787e-5}{\GeV^{-2}} .
	\nonumber
\end{align}
The running of the strong coupling ($\alphas$) was evaluated using the LHAPDF library with the associated PDF set, while the \MSbar mass of the bottom quark ($\overline{m}_{\Pqb}$) was directly computed within \NNLOJET.
Finally, in the case of $\PWpm\PH$ production, we assumed a diagonal CKM matrix for the vector-boson--quark couplings.

For the unphysical scales appearing in the calculation, we chose to set and vary them independently for the production and decay parts.
The central factorisation and renormalisation scales of the production sub-processes were chosen to the invariant mass of the $V\PH$ system~$M_{V\PH}$, whereas the central renormalisation scale of the decay was set to the Higgs-boson mass~$m_{\PH}$.
We evaluate the differential cross section for a total of 21 different scale settings that are obtained from all possible combinations of
\begin{align}
	\muF              & = M_{V\PH} \, \left[ 1, \tfrac{1}{2}, 2 \right], &
	\muR^\text{prod.} & = M_{V\PH} \, \left[ 1, \tfrac{1}{2}, 2 \right], &
	\muR^\text{dec.}  & = m_{\PH}  \, \left[ 1, \tfrac{1}{2}, 2 \right],
	\label{eq:scales}
\end{align}
with the additional constraint $\tfrac{1}{2}\leq \muF/\muR^\text{prod.} \leq 2$ following the conventional 7-point scale variation for the production sub-process.

\subsection{Fiducial cross section}
\label{sec:fiducial}

\begin{table}
	\centering
	\begin{tabular}{r @{\qquad} c c c}
		\toprule
		                                    & $\PWp\PH $                                  & $\PWm\PH $                                  & $ \PZ\PH $                                 \\
		\midrule
		$\sigma_{\text{LO}} \, [\si{\fb}]$  & $\num{18.06}\,^{+\num{2.87}}_{-\num{2.41}}$ & $\num{11.96}\,^{+\num{1.90}}_{-\num{1.60}}$ & $\num{4.83}\,^{+\num{0.77}}_{-\num{0.65}}$ \\
		$\sigma_{\text{NLO}} \, [\si{\fb}]$ & $\num{21.52}\,^{+\num{0.88}}_{-\num{1.08}}$ & $\num{14.21}\,^{+\num{0.58}}_{-\num{0.71}}$ & $\num{5.71}\,^{+\num{0.22}}_{-\num{0.28}}$ \\
		$\sigma_{\text{NNLO}}\, [\si{\fb}]$ & $\num{20.68}\,^{+\num{0.16}}_{-\num{0.46}}$ & $\num{13.64}\,^{+\num{0.11}}_{-\num{0.31}}$ & $\num{5.92}\,^{+\num{0.13}}_{-\num{0.16}}$ \\
		\bottomrule
	\end{tabular}
	\caption{The fiducial cross sections for all $V\PH$ processes according to the setup of section~\ref{sec:setup}.
		The error on the values represents the theoretical uncertainty which was obtained by taking the minimum and maximum values of the 21-point scale variation.}
	\label{tab:fid}
\end{table}

The cross-section predictions including fiducial cuts for the different $V\PH$ processes are summarised in table~\ref{tab:fid} at the various orders in \alphas.

Regarding the $\PWpm\PH$ fiducial values, we observe an $\cO(20\%)$ increase in the cross section from the NLO corrections and a slight $\cO(5\%)$ decrease when going from NLO to NNLO.
The minimum and the maximum values of the 21-point scale variations yield the theoretical uncertainties of our predictions, which are $\cO(15\%)$ at LO, $\cO(5\%)$ at NLO, and reduce to only $\cO(2\%)$ at NNLO with a three-fold asymmetry between the lower and upper bounds of the latter values.
The decrease in the size of the theoretical uncertainty is apparent at each of these orders, demonstrating the perturbative convergence of these results in a satisfying manner.
This will be further accentuated for flavour-sensitive jet observables in section~\ref{sec:results_dist}.

For the $\PZ\PH$ fiducial values we see a different behaviour beyond NLO: the gluon--gluon-induced $\PZ\PH$-only top loop contributions of figures~\ref{fig:top-self-zh-i} and~\ref{fig:top-self-zh-ii} dominate the NNLO coefficient such that there is an $\cO(4\%)$ increase going from NLO to NNLO, contrasting the decrease seen for the $\PWpm\PH$ case.
The $\PZ\PH$-exclusive channels open up at NNLO, and therefore the theoretical uncertainty does not exhibit such a strong reduction in size but remains around~$\cO(3\%)$.

Note that the reduction of scale uncertainties observed here is spoiled in all cases when a rescaling prescription is employed that incorporates a fixed branching ratio for the $\PH\to\Pqb\Paqb$ decay, as is commonly done in previous calculations for the $V\PH$ processes.
A comparison of our results in table~\ref{tab:fid} to such a rescaled cross-section prediction is presented in appendix~\ref{sec:fixbr}.

\subsection{Scale variations}
\label{sec:scale}

The dependence on the renormalisation scales $\muR^{\text{prod.}/\text{dec.}}$ can serve as a non-trivial check of the final results obtained from the numerical computation.
To this end, we ensure that the different scale settings of eq.~\eqref{eq:scales} are correctly reproduced by the analytic renormalisation-group running starting from the central scale choice.%
\footnote{
	For processes involving just a production part, the analytic expressions have been explicitly given in ref.~\cite{Currie:2018xkj}.
}
This is of particular importance for the calculation at hand, as the independent variation of scales for the different sub-processes was for the first time implemented in the \NNLOJET framework for the present work.

\begin{figure}
	\centering
	\renewcommand\thesubfigure{a}
	\begin{subfigure}[h]{.48\textwidth}
		\includegraphics[width=\textwidth,page=1]{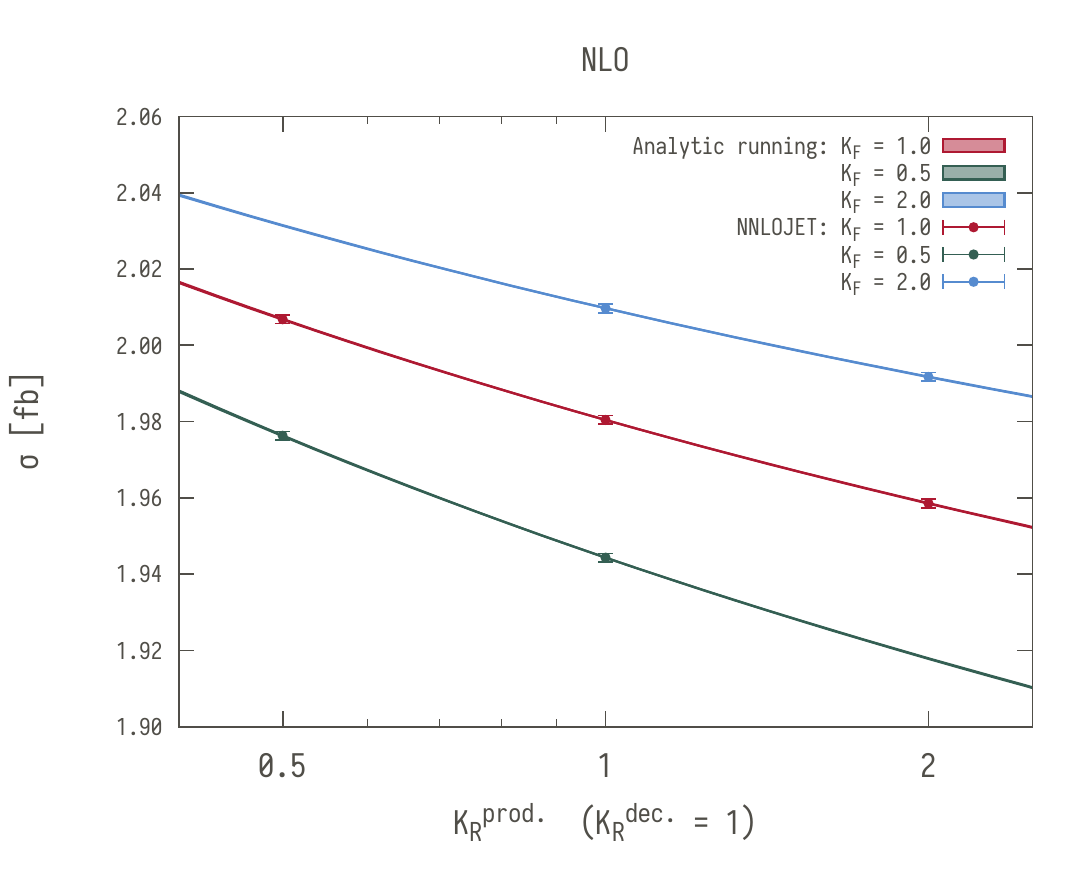}
		\subcaption{}
		\label{fig:scales_a}
	\end{subfigure}
	\hfill
	\renewcommand\thesubfigure{d}
	\begin{subfigure}[h]{.48\textwidth}
		\includegraphics[width=\textwidth,page=4]{{var_plot_paper}.pdf}
		\subcaption{}
		\label{fig:scales_d}
	\end{subfigure}
	\newline
	\renewcommand\thesubfigure{b}
	\begin{subfigure}[h]{.48\textwidth}
		\includegraphics[width=\textwidth,page=2]{{var_plot_paper}.pdf}
		\subcaption{}
		\label{fig:scales_b}
	\end{subfigure}
	\hfill
	\renewcommand\thesubfigure{e}
	\begin{subfigure}[h]{.48\textwidth}
		\includegraphics[width=\textwidth,page=5]{{var_plot_paper}.pdf}
		\subcaption{}
		\label{fig:scales_e}
	\end{subfigure}
	\newline
	\renewcommand\thesubfigure{c}
	\begin{subfigure}[h]{.48\textwidth}
		\includegraphics[width=\textwidth,page=3]{{var_plot_paper}.pdf}
		\subcaption{}
		\label{fig:scales_c}
	\end{subfigure}
	\hfill
	\renewcommand\thesubfigure{f}
	\begin{subfigure}[h]{.48\textwidth}
		\includegraphics[width=\textwidth,page=6]{{var_plot_paper}.pdf}
		\subcaption{}
		\label{fig:scales_f}
	\end{subfigure}
	\caption{Numerical versus analytical scale variation of the $\PWp\PH$ process at NLO~(left) and NNLO~(right) for the bin $\SI{220}{\GeV} \le M_{\PW\PH} \le \SI{230}{\GeV}$ and three different slices in the $(\muR^\text{prod.},\muR^\text{dec.})$ plane.
	}
	\label{fig:scales}
\end{figure}

The comparison between the analytic evolution and the 21 points obtained from the numerical computation using \NNLOJET are shown in figure~\ref{fig:scales} for the case of the $\PWp\PH$ process at NLO~(\subref{fig:scales_a}--\subref{fig:scales_c}) and NNLO~(\subref{fig:scales_d}--\subref{fig:scales_f}).
We performed a scan in the two-dimensional $(\muR^\text{prod.},\muR^\text{dec.})$ space by choosing three different slices that cover the combinations in eq.~\eqref{eq:scales} where the three choices in the factorisation scale $\muF=M_{\PW\PH}  \;\left[ 1, \tfrac{1}{2}, 2 \right ]$ are shown as separate curves:
\begin{itemize}
	\item[(a,d)] We keep the decay renormalisation scale fixed to $\muR^{\text{dec.}} = m_{\PH}$ and vary the scale in the production sub-process according to
	      \begin{align}
		      \muR^{\text{prod.}} & = K_\mathrm{R}^{\text{prod.}} \times M_{\PW\PH} & \text{with } K_\mathrm{R}^{\text{prod.}} \in \left[ \tfrac{1}{2}, 2 \right ].
	      \end{align}
	\item[(b,e)] We keep the production renormalisation scale fixed to $\muR^{\text{prod.}} = M_{\PW\PH}$ and vary the scale in the decay sub-process according to
	      \begin{align}
		      \muR^{\text{dec.}} & = K_\mathrm{R}^{\text{dec.}} \times m_{\PH} & \text{with } K_\mathrm{R}^{\text{dec.}} \in \left[ \tfrac{1}{2}, 2 \right ].
	      \end{align}
	\item[(c,f)] We choose a diagonal slice in the $(\muR^\text{prod.},\muR^\text{dec.})$ plane setting $K_\mathrm{R}^{\text{prod.}} = K_\mathrm{R}^{\text{dec.}} \equiv K_\mathrm{R}$ with the individual scales given as
	      \begin{align}
		      \muR^{\text{prod.}} & = K_\mathrm{R} \times M_{\PW\PH} , &
		      \muR^{\text{dec.}}  & = K_\mathrm{R} \times m_{\PH}      &
		      \text{with } K_\mathrm{R} \in \left[ \tfrac{1}{2}, 2 \right ].
	      \end{align}
\end{itemize}
Note that the invariant mass $M_{\PW\PH}$ constitutes a dynamical quantity that varies on an event-by-event basis.
The curves in figure~\ref{fig:scales} are obtained by picking a specific bin $M_{\PW\PH} \in [220,230]~\GeV$ to assign a value to the production scale, where the width of the bands in the smooth curves correspond to the uncertainty that arises from the finite bin width.

We observe an excellent agreement between the numerical results from \NNLOJET and the curves predicted from the renormalisation group equations.
The dramatic reduction in scale uncertainties can be further appreciated by contrasting the vertical ranges in the figures at NLO~(left) and NNLO~(right).
We carried out the same tests also for the $\PWm\PH$ and the $\PZ\PH$ processes as well as for other individual $M_{V\PH}$ bins in the distributions and found that the scale variation of the numerical results match the analytical formul\ae\ in all cases.

\subsection{Distributions}
\label{sec:results_dist}

\begin{figure}
	\centering
	\begin{subfigure}[h]{.48\textwidth}
		\includegraphics[width=\textwidth]{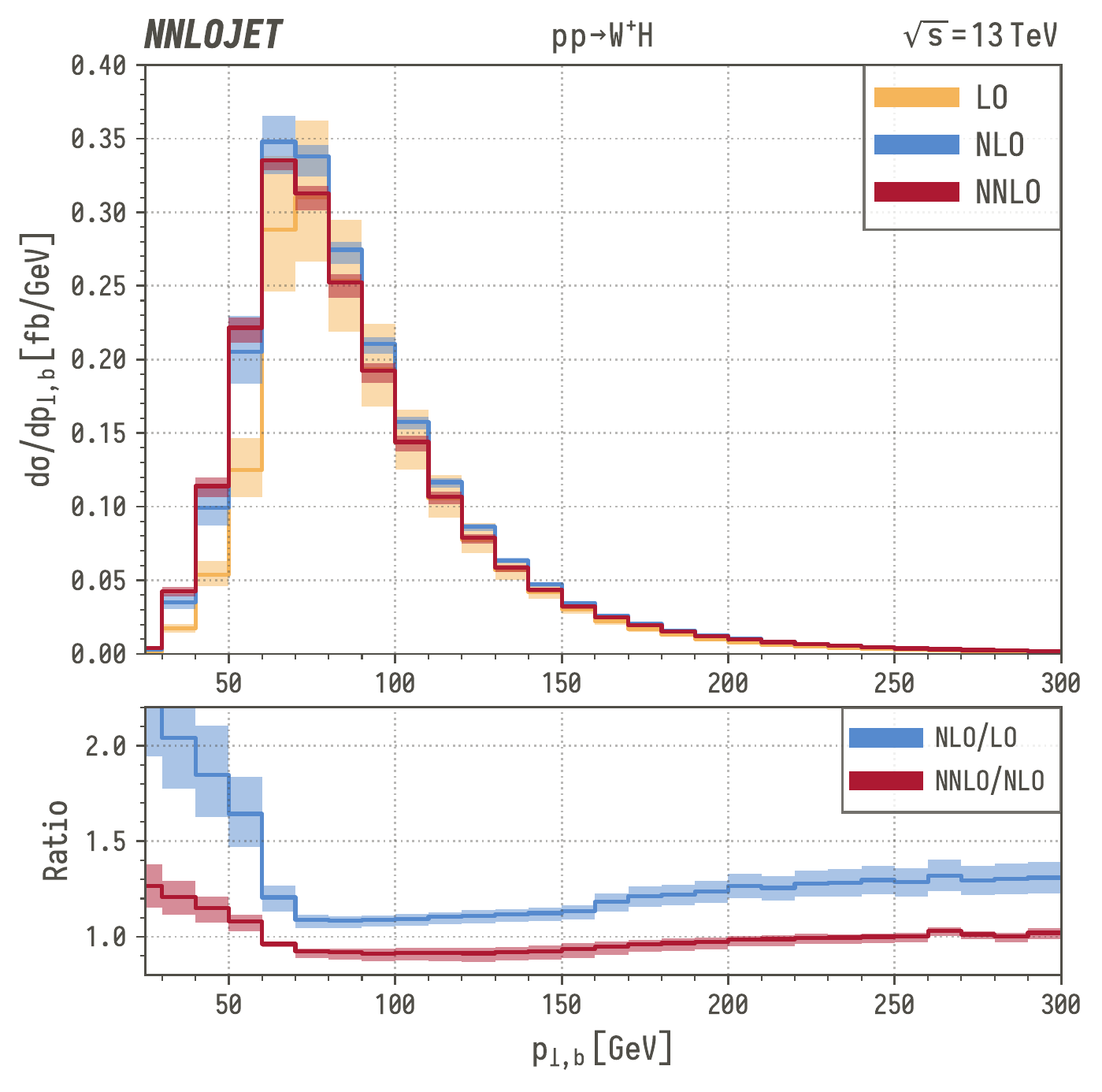}
		\subcaption{}
		\label{fig:wph_ptb1}
	\end{subfigure}
	\begin{subfigure}[h]{.48\textwidth}
		\includegraphics[width=\textwidth]{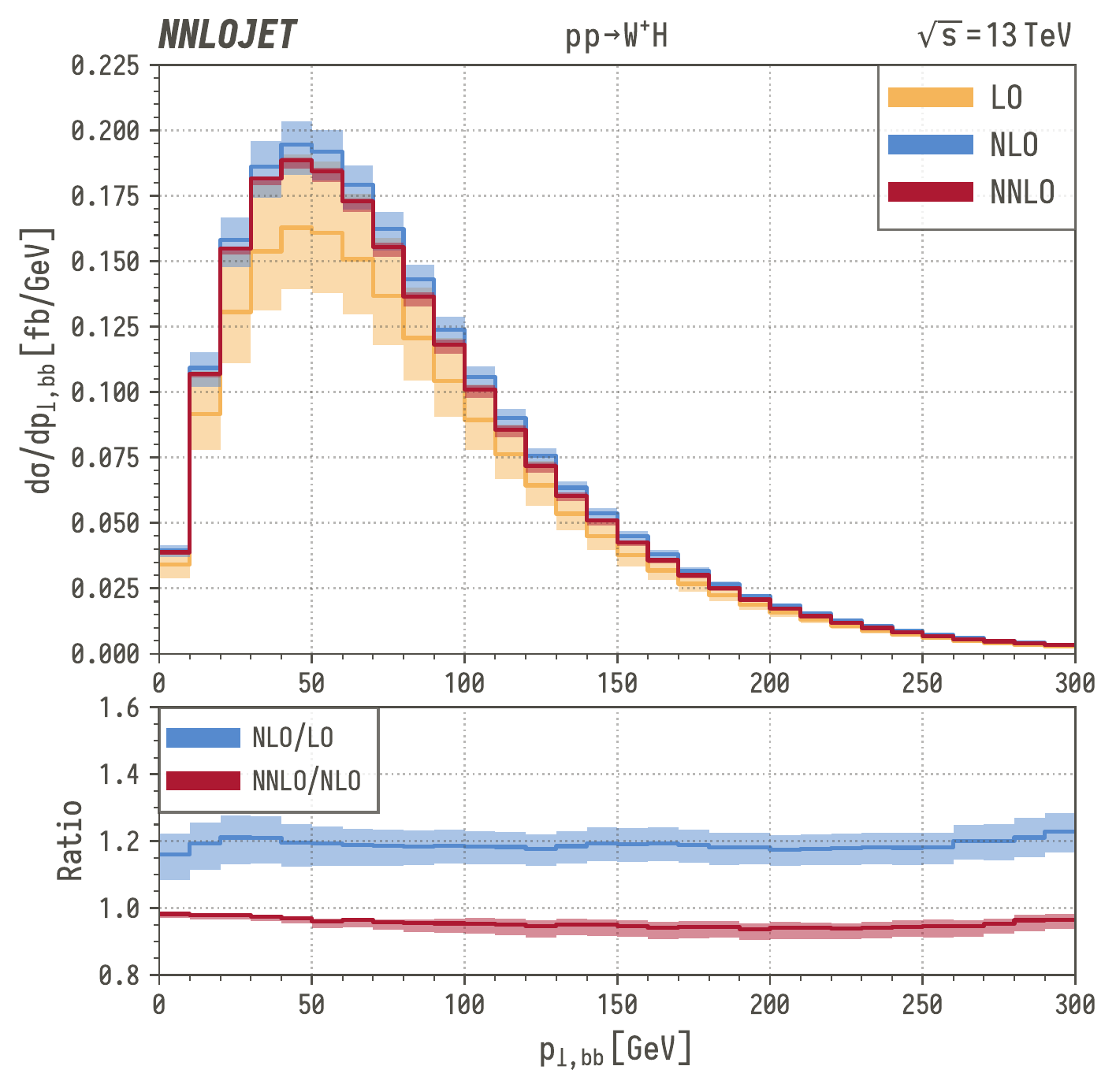}
		\subcaption{}
		\label{fig:wph_ptbb}
	\end{subfigure}

	\vspace{3em}

	\begin{subfigure}[h]{.48\textwidth}
		\includegraphics[width=\textwidth]{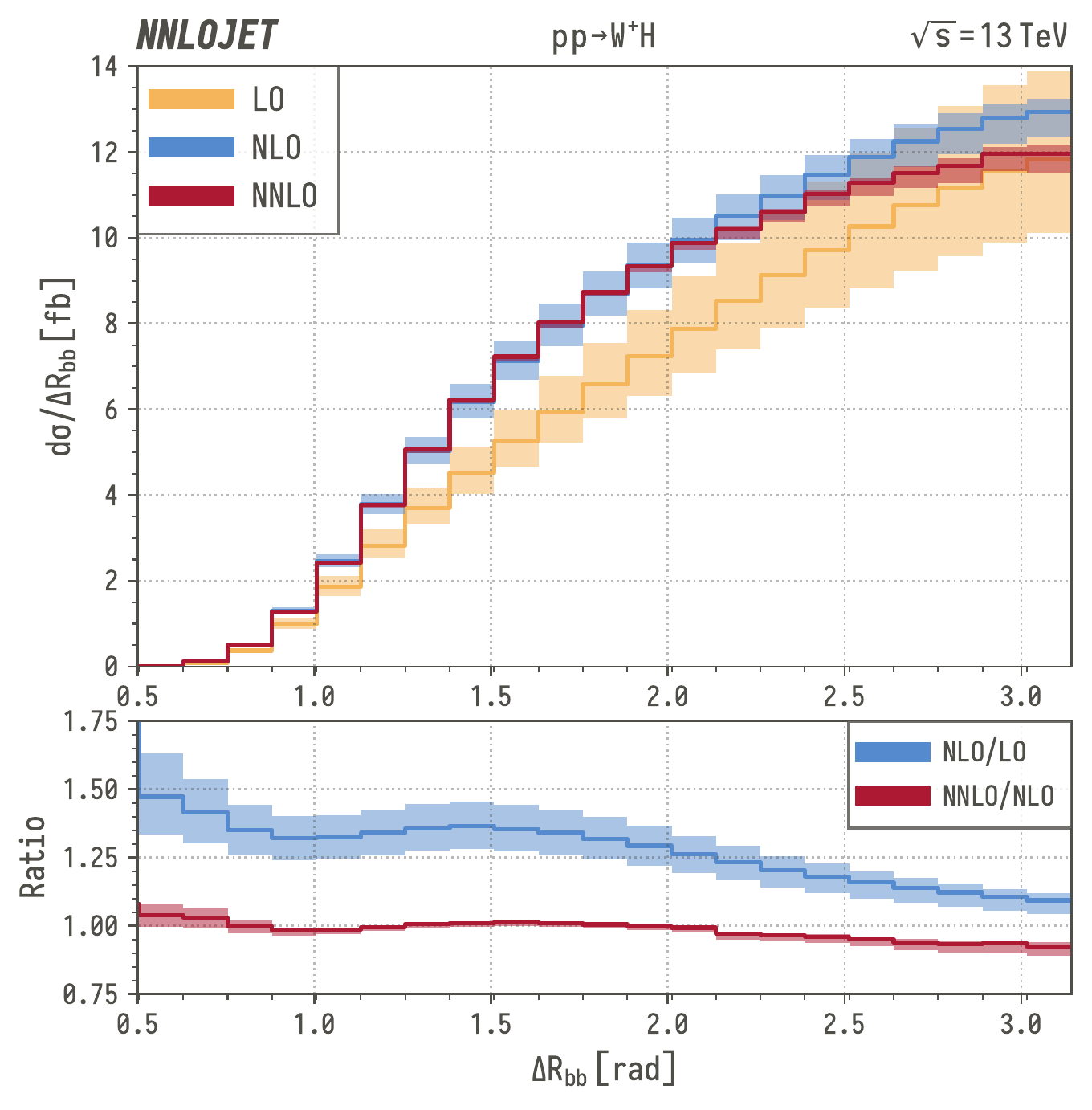}
		\subcaption{}
		\label{fig:wph_dRbb}
	\end{subfigure}
	\begin{subfigure}[h]{.48\textwidth}
		\includegraphics[width=\textwidth]{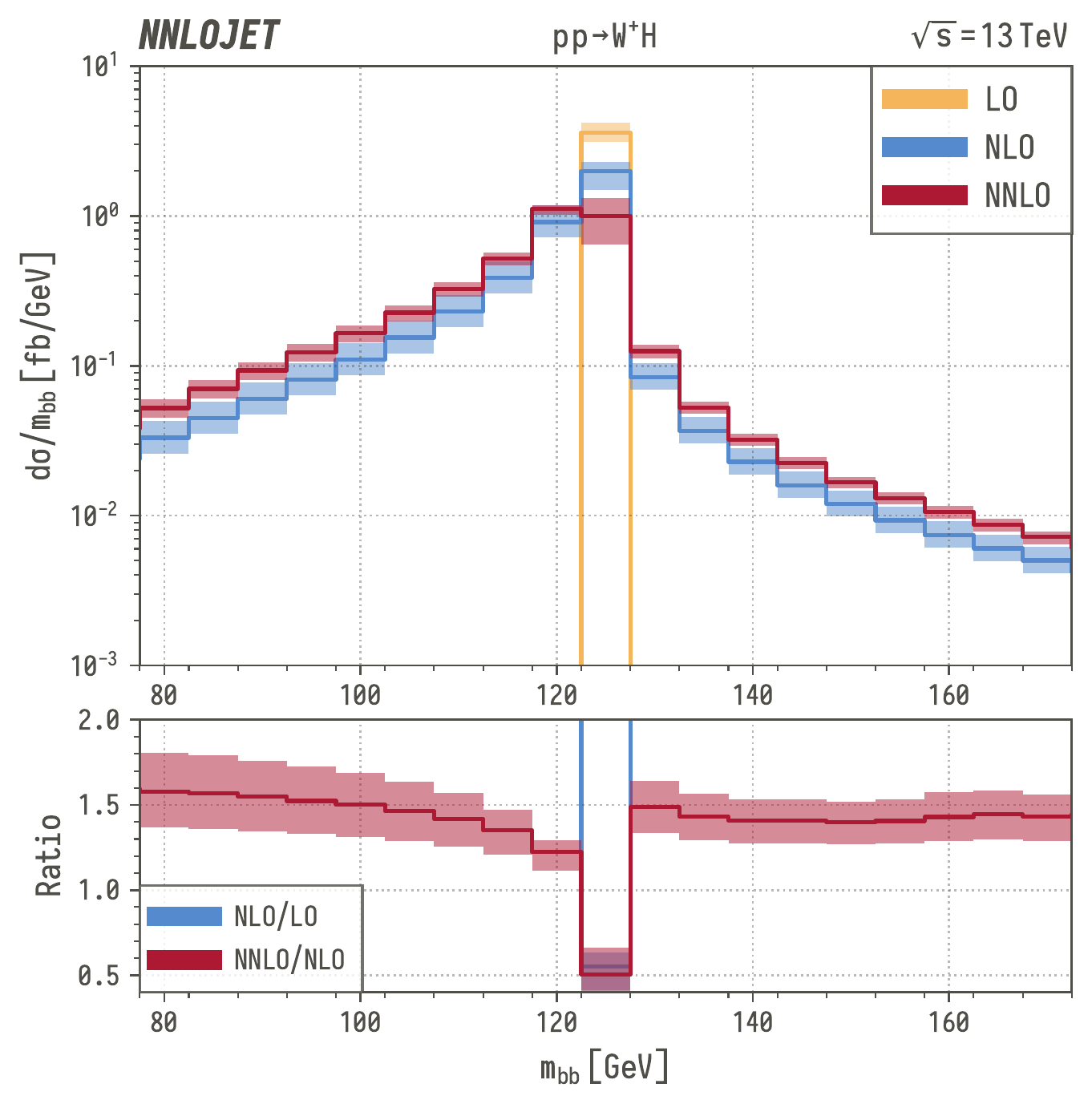}
		\subcaption{}
		\label{fig:wph_mbb}
	\end{subfigure}
	\caption{Flavour-sensitive jet distributions for the $\PWp\PH$ process showing (\subref{fig:wph_ptb1})~the transverse momentum of the leading \Pqb-jet, (\subref{fig:wph_ptbb})~the transverse momentum of the \Pqb-jet pair, (\subref{fig:wph_dRbb})~the angular separation of the \Pqb-jet pair, and~(\subref{fig:wph_mbb})~the invariant mass of the \Pqb-jet pair closest to the Higgs boson mass.
		The upper panel contains the absolute values while the lower panel shows the bin-by-bin ratios with respect to the previous order evaluated at the central scale.}
	\label{fig:wph_dist}
\end{figure}

\begin{figure}
	\centering
	\begin{subfigure}[h]{.48\textwidth}
		\includegraphics[width=\textwidth]{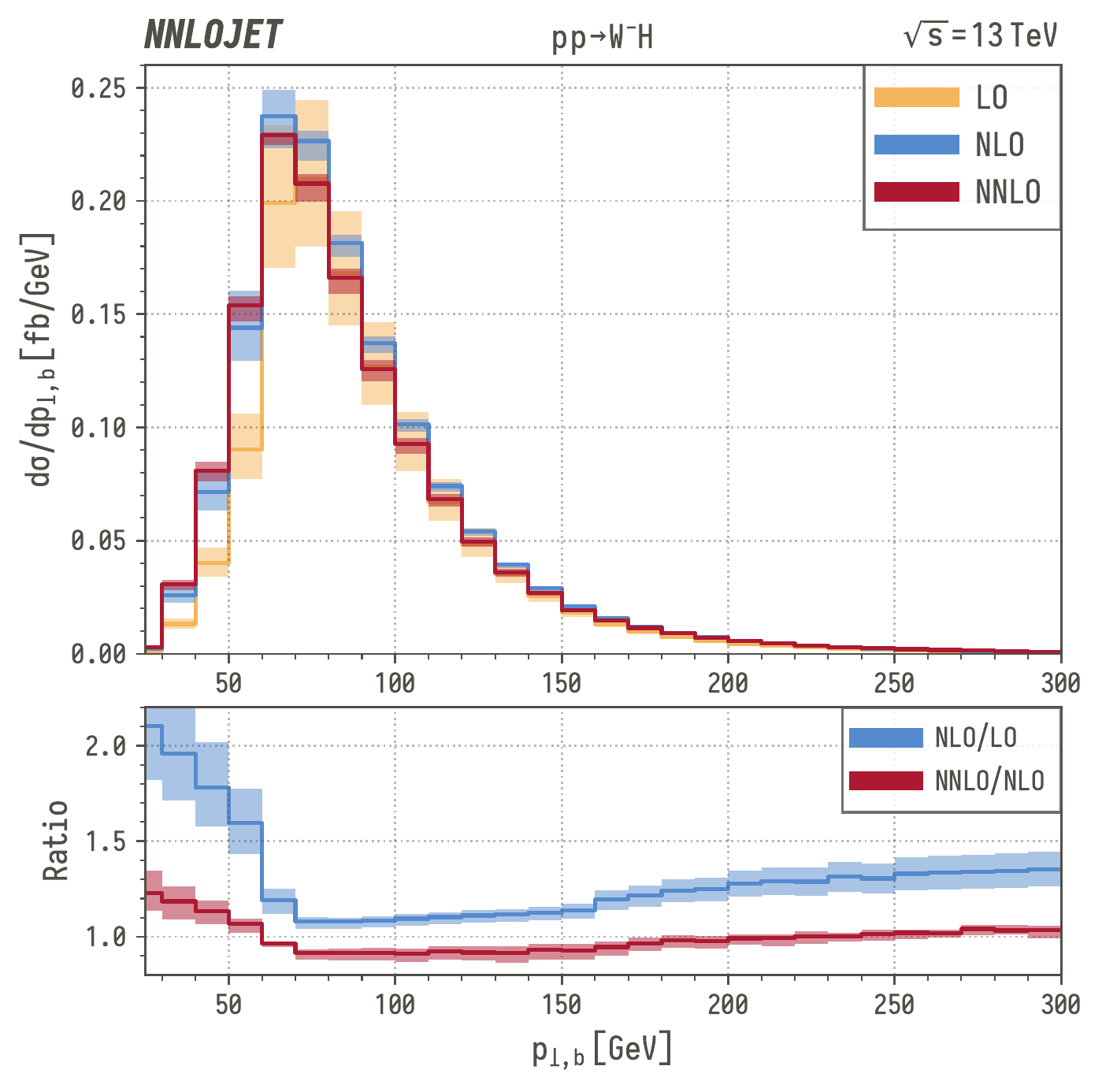}
		\subcaption{}
		\label{fig:wmh_ptb1}
	\end{subfigure}
	\begin{subfigure}[h]{.48\textwidth}
		\includegraphics[width=\textwidth]{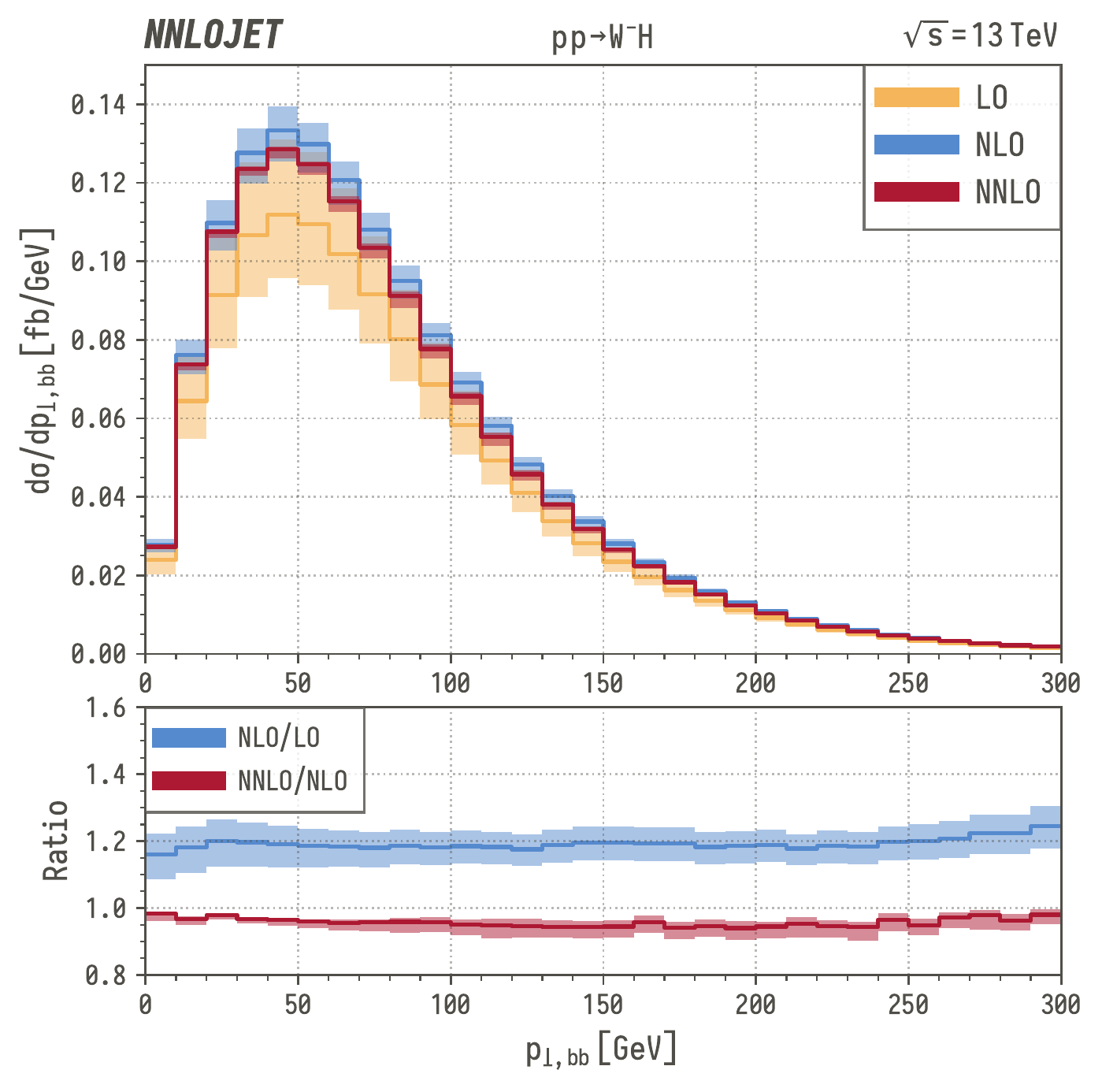}
		\subcaption{}
		\label{fig:wmh_ptbb}
	\end{subfigure}

	\vspace{3em}

	\begin{subfigure}[h]{.48\textwidth}
		\includegraphics[width=\textwidth]{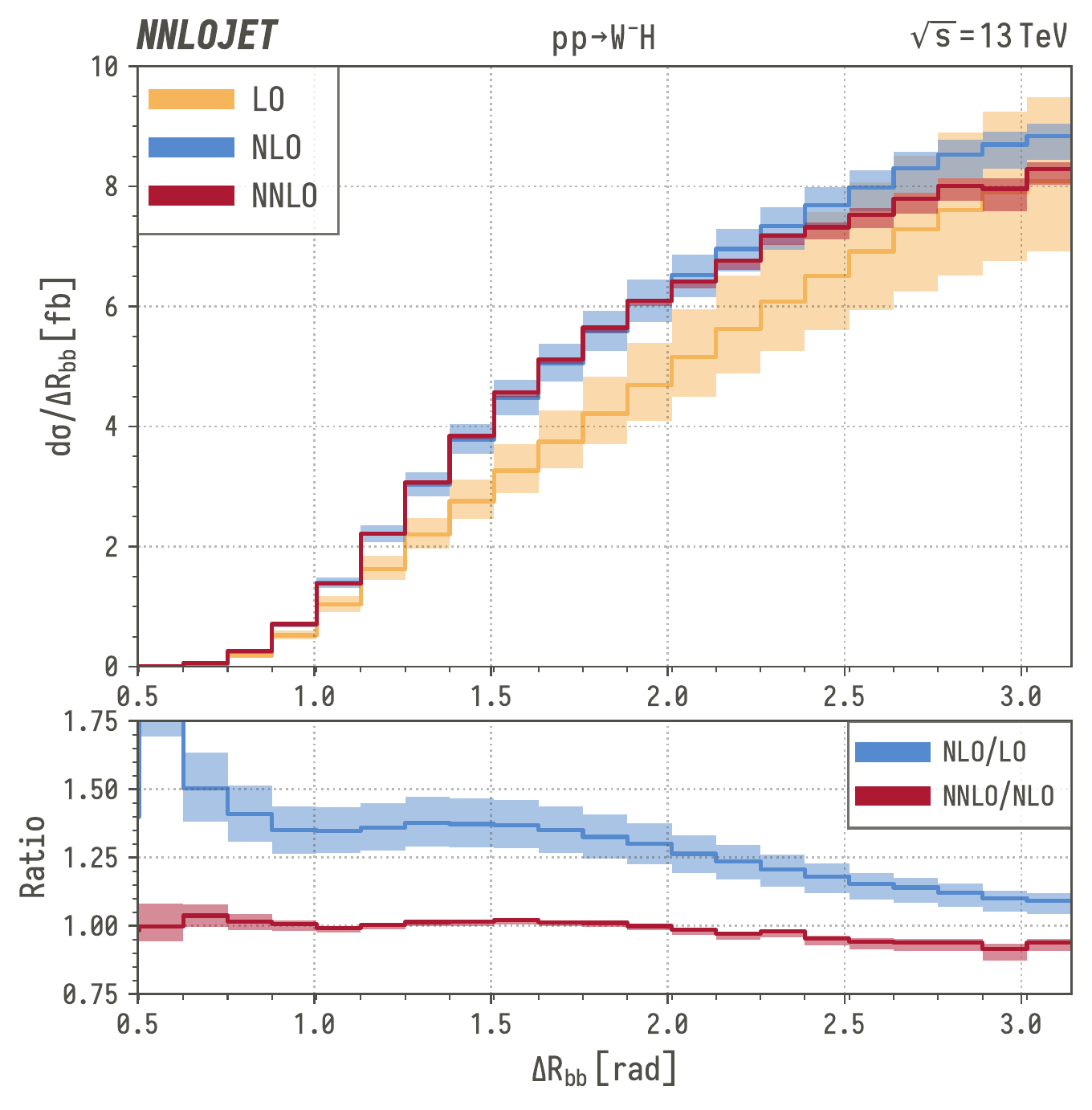}
		\subcaption{}
		\label{fig:wmh_dRbb}
	\end{subfigure}
	\begin{subfigure}[h]{.48\textwidth}
		\includegraphics[width=\textwidth]{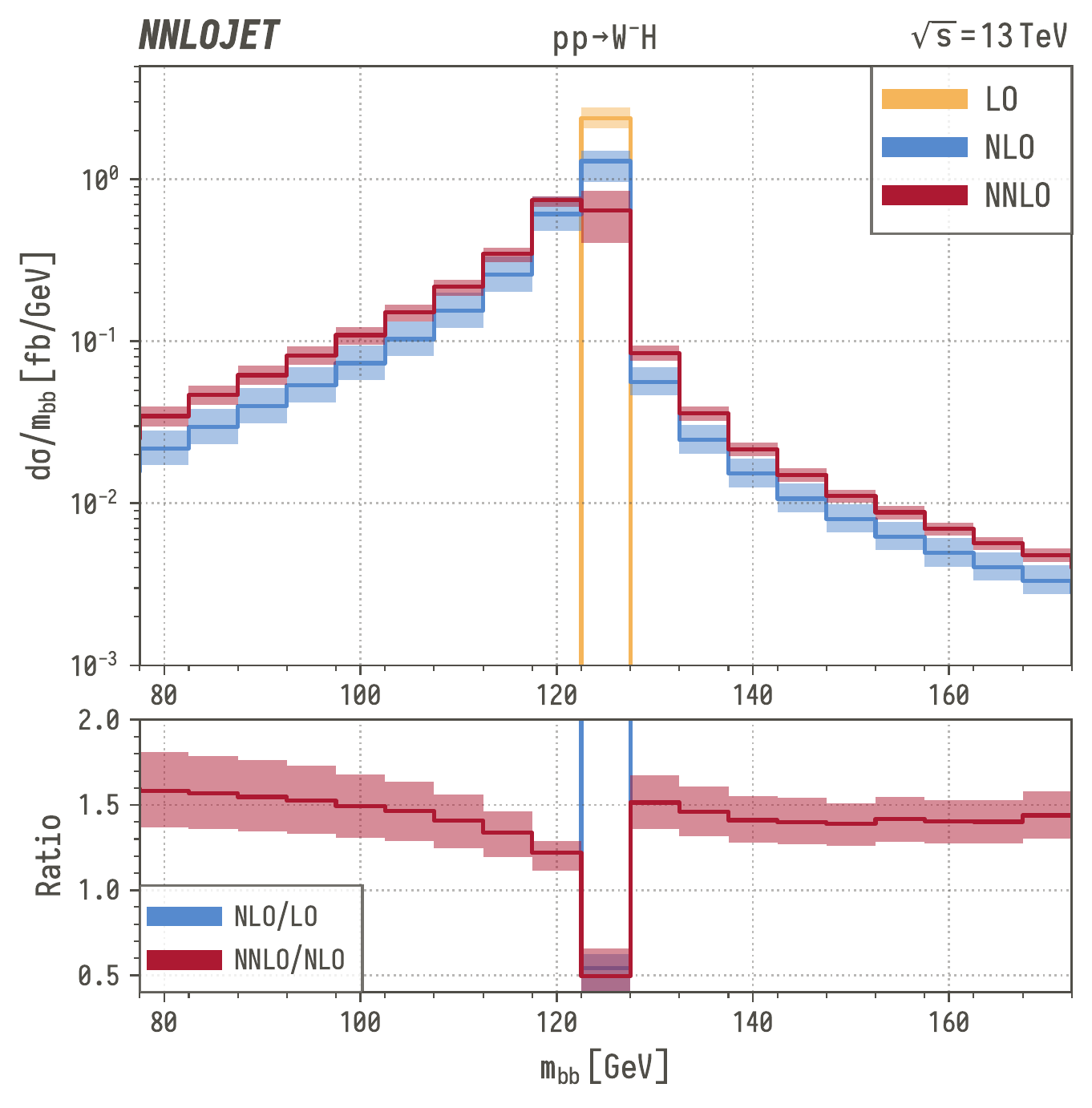}
		\subcaption{}
		\label{fig:wmh_mbb}
	\end{subfigure}
	\caption{Flavour-sensitive jet distributions for the $\PWm\PH$ process showing (\subref{fig:wmh_ptb1})~the transverse momentum of the leading \Pqb-jet, (\subref{fig:wmh_ptbb})~the transverse momentum of the \Pqb-jet pair, (\subref{fig:wmh_dRbb})~the angular separation of the \Pqb-jet pair, and~(\subref{fig:wmh_mbb})~the invariant mass of the \Pqb-jet pair closest to the Higgs boson mass.
		The upper panel contains the absolute values while the lower panel shows the bin-by-bin ratios with respect to the previous order evaluated at the central scale.}
	\label{fig:wmh_dist}
\end{figure}

\begin{figure}
	\centering
	\begin{subfigure}[h]{.48\textwidth}
		\includegraphics[width=\textwidth]{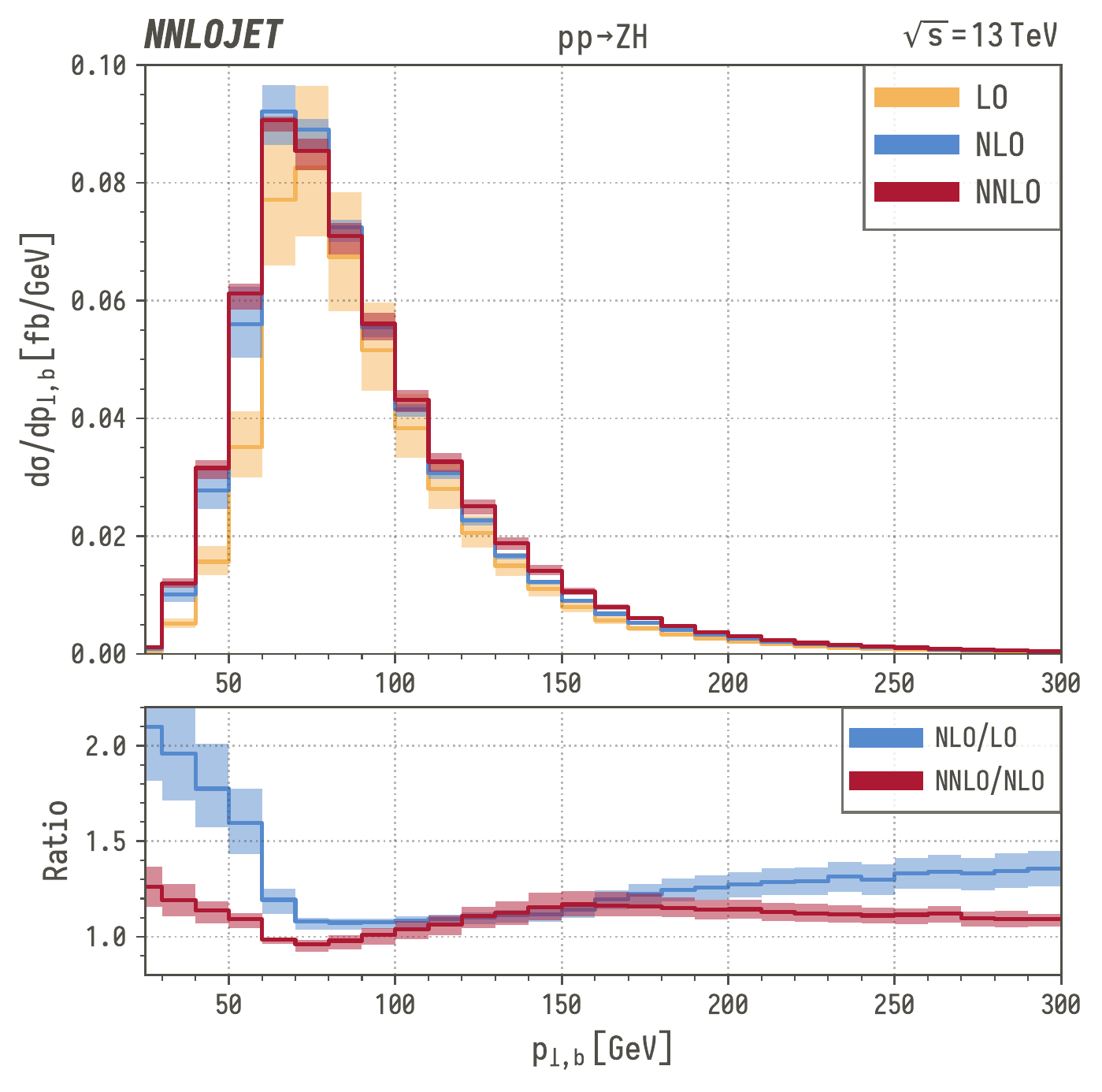}
		\subcaption{}
		\label{fig:zh_ptb1}
	\end{subfigure}
	\begin{subfigure}[h]{.48\textwidth}
		\includegraphics[width=\textwidth]{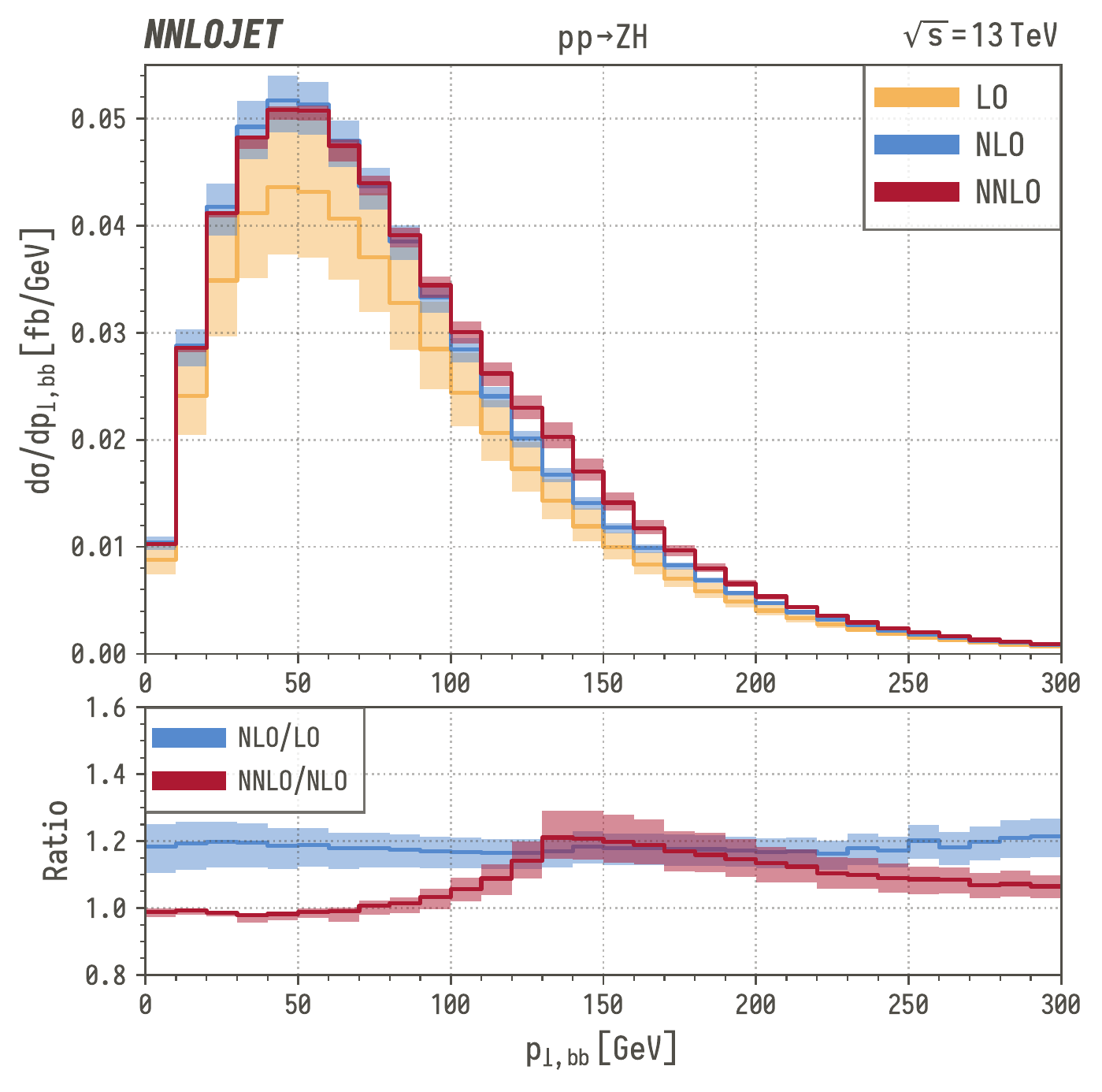}
		\subcaption{}
		\label{fig:zh_ptbb}
	\end{subfigure}

	\vspace{3em}

	\begin{subfigure}[h]{.48\textwidth}
		\includegraphics[width=\textwidth]{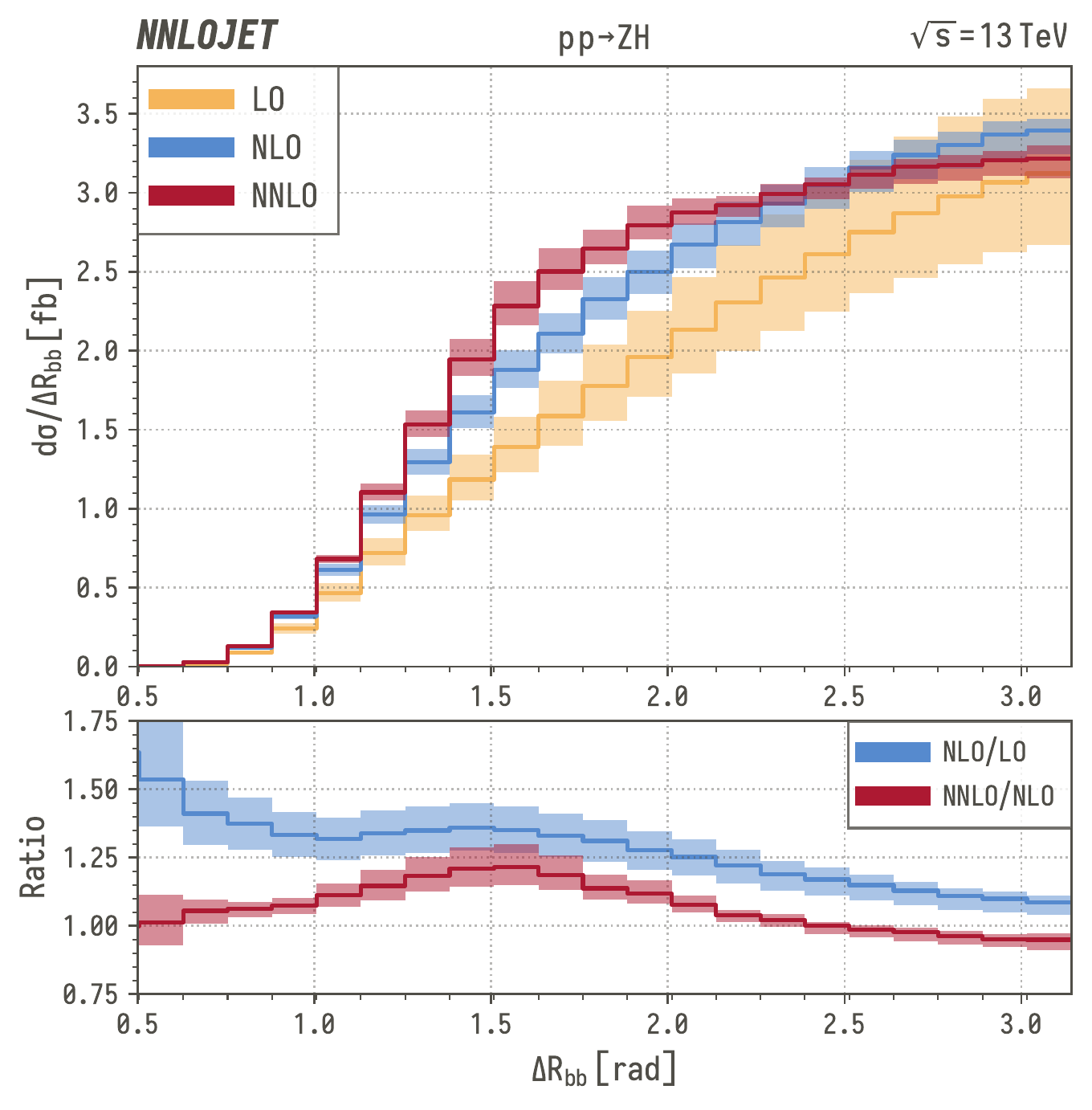}
		\subcaption{}
		\label{fig:zh_dRbb}
	\end{subfigure}
	\begin{subfigure}[h]{.48\textwidth}
		\includegraphics[width=\textwidth]{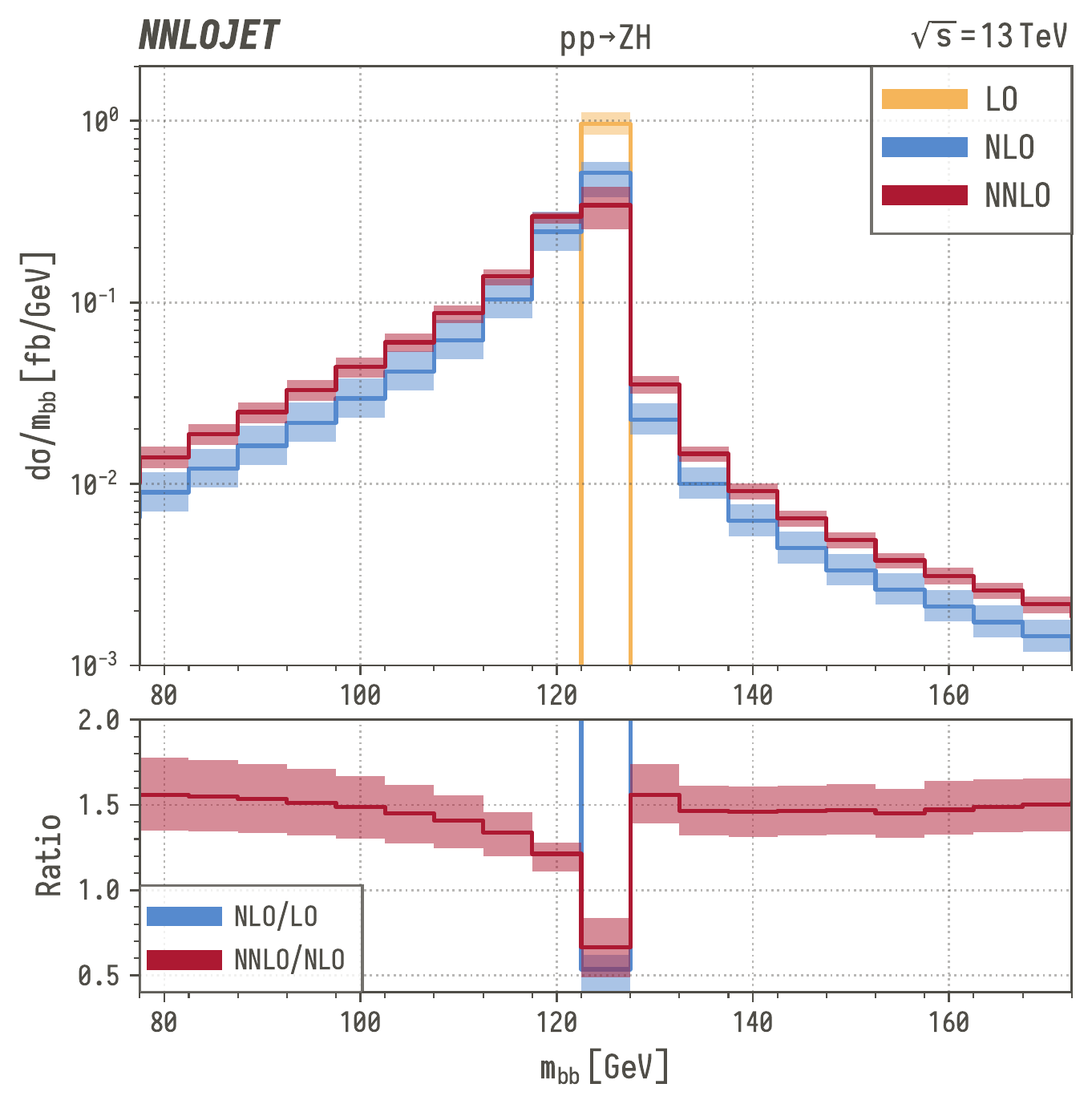}
		\subcaption{}
		\label{fig:zh_mbb}
	\end{subfigure}
	\caption{Flavour-sensitive jet distributions for the $\PZ\PH$ process showing (\subref{fig:zh_ptb1})~the transverse momentum of the leading \Pqb-jet, (\subref{fig:zh_ptbb})~the transverse momentum of the \Pqb-jet pair, (\subref{fig:zh_dRbb})~the angular separation of the \Pqb-jet pair, and~(\subref{fig:zh_mbb})~the invariant mass of the \Pqb-jet pair closest to the Higgs boson mass.
		The upper panel contains the absolute values while the lower panel shows the bin-by-bin ratios with respect to the previous order evaluated at the central scale.}
	\label{fig:zh_dist}
\end{figure}

In figures~\ref{fig:wph_dist}--\ref{fig:zh_dist} we present differential distributions of flavour-sensitive observables for the three different associated Higgs boson production processes $\PWp\PH$, $\PWm\PH$, and $\PZ\PH$:%
\footnote{
	We focus on this set of observables in order to allow for a qualitative comparison with refs.~\cite{Ferrera:2017zex,Caola:2017xuq}.
}
\begin{enumerate}[label=(\alph*)]
	\item the transverse momentum $p_{\bot,\Pqb}$ of the leading \Pqb-jet,
	\item the transverse momentum $p_{\bot,\Pqb\Pqb}$ of the pair of two \Pqb-jets,
	\item the angular separation $\Delta R_{\Pqb\Pqb} = \sqrt{\Delta \eta_{\Pqb\Pqb}^2 + \Delta \phi_{\Pqb\Pqb}^2}$ of two \Pqb-jets,
	\item and the invariant mass $m_{\Pqb\Pqb}$ of two \Pqb-jets,
\end{enumerate}
where in~(b--d) the two \Pqb-jets are selected whose invariant mass is closest to $m_\PH$ in order to identify the candidate pair that is most likely to originate from the Higgs decay.

Up to NLO, all three production modes of $\PWp\PH$, $\PWm\PH$, and $\PZ\PH$ show similar qualitative behaviour for all four investigated distributions.
However, there are significant phenomenological differences between the $\PWpm\PH$ and $\PZ\PH$ distributions at NNLO.

NNLO corrections to the $\PWpm\PH$ cases lead to substantial stabilisation of the predictions for the first three distributions shown in figures~\ref{fig:wph_dist}--\ref{fig:wmh_dist}, parts~(\subref{fig:wph_ptb1}--\subref{fig:wph_dRbb}): size and shape are only slightly modified at NNLO compared to the NLO predictions; the scale-variation bands, however, are reduced considerably.
In contrast, the first three of the $\PZ\PH$ distributions show an excess of events in the central regions throughout figure~\ref{fig:zh_dist}, parts~(\subref{fig:zh_ptb1}--\subref{fig:zh_dRbb}).
This behaviour is attributed to top-quark-loop threshold effects in the dominant gluon--gluon-induced $\PZ\PH$-exclusive amplitudes of figures~\ref{fig:top-self-zh-i} and~\ref{fig:top-self-zh-ii}.
As mentioned earlier, these channels first contributed at NNLO, which also explains the widening of the theoretical uncertainty bands around the threshold regions of these distributions.

Concerning the invariant mass distribution of all three production modes shown in figures~\ref{fig:wph_mbb}--\ref{fig:zh_mbb}, the features previously noted in refs.~\cite{Ferrera:2017zex,Caola:2017xuq} can be confirmed by our predictions as well:
due to the very narrow width of the Higgs boson, the $m_{\Pqb\Pqb}$ distribution has a natural kinematic threshold at $m_{\PH}=\SI{125.09}{\GeV}$ and the phase space away from this value is barely populated at LO.
Consequently, NNLO corrections are effectively NLO-accurate for most of the bins, which explains the large corrections and relatively larger uncertainty bands for these distributions.
The left shoulder below $m_{\PH}$ is mainly the result of radiation from the decay, whereas the shoulder above $m_{\PH}$ is due to radiative corrections to the production.
Fixed-order predictions at the threshold region of $m_{\Pqb\Pqb} \approx m_{\PH}$, however, should not be trusted as they are prone to Sudakov-type instabilities.
A proper treatment of this region would require the inclusion of resummation effects.
In our case, the binning is sufficiently coarse so that such instabilities only manifest in larger uncertainty bands for the $m_{\Pqb\Pqb} = m_{\PH}$ bin and not as an explicit divergence.

%% file: VH-sec5.tex
\section{Summary and conclusions}
\label{sec:summary}

We reported on the calculation of NNLO corrections to the Higgs Strahlung processes $\PWp\PH$, $\PWm\PH$, and $\PZ\PH$ including the off-shell leptonic decay of the gauge boson as well as the Higgs decaying into a bottom--antibottom pair.
The calculation consistently takes into account NNLO corrections to the production and decay sub-processes and fully retains the differential information on the final state.

The study of $V\PH$ ($\PH\to\Pqb\Paqb$) processes critically relies on the tagging of bottom jets in order to isolate the candidate pairs associated to the Higgs boson.
We described our independent implementation of the infrared-safe flavour-$k_t$ algorithm in the \NNLOJET parton-level event generator and the necessary modifications this entails in the framework of the antenna subtraction formalism.

A detailed account was given on the residual theory uncertainties by allowing the scales in the production and decay sub-processes to vary independently.
This conservative approach resulted into taking the envelope of 21 scale variations for the full process but allowed for a more comprehensive view into the impact of higher orders on the reduction of scale uncertainties.
The NNLO corrections to the fiducial cross section were found to exhibit a good perturbative convergence with residual uncertainties at the percent level.
We contrasted our naïve perturbative expansion of the cross section with a more commonly employed rescaling procedure using the branching ratio $\mathrm{BR}(\PH\to\Pqb\Paqb)$, where we observed the latter to overestimate the residual scale uncertainties.
This was attributed to a miscancellation in the scale dependence among the terms that receive different rescaling factors and lead us to advocate the simpler prescription to be more reliable beyond NLO.

Flavour-sensitive observables were studied by investigating differential distributions where a similar stabilisation of the perturbative series was found as in the cross sections.
Larger effects from higher-order corrections were seen in the invariant mass distributions of two \Pqb-jets, which can be attributed to this observables being only NLO-accurate away from $m_{\Pqb\Pqb}\sim m_\PH$.
A comparison between the $\PWpm\PH$ and $\PZ\PH$ processes showed a qualitatively similar behaviour but also emphasised the phenomenologically sizeable impact that arise from the gluon--gluon-induced top-quark loop amplitudes.

The study of flavour-sensitive jet observables with fixed-order predictions, such as those associated to \Pqb-jets in the present work, must be performed in an infrared-safe way. For calculations based on massless QCD this can only be achieved with a flavour-aware jet algorithm (such as flavour-$k_t$), while for a massive calculation this is achievable with a flavour-blind algorithm (such as anti-$k_t$).
In many cases the corresponding massive calculation may not be available, or the massless calculation may actually be preferred due to the presence of large logarithmic corrections which may be easily resummed via PDF evolution.
Future comparisons to measurements are only viable if a similar prescription is also employed in the experiment, and the application of the even-tag exclusion here was mainly motivated to facilitate the experimental implementation.
The use of flavour-sensitive jet algorithms is not only important to the $V\PH$ process class but we expect it to be of relevance for \emph{any flavour-sensitive jet observable}, such as the associated production of the flavoured jet with a gauge boson.
Such studies will be left for future work.

%% file: VH-AppendixA.tex
\section{Effects of even-tag exclusion}
\label{sec:eventag}

\begin{table}
	\centering
	\sisetup{
		table-figures-uncertainty = 1,
		table-number-alignment = center,
		separate-uncertainty
	}
	\begin{tabular}{ r c }
		\toprule
		\multicolumn{1}{c}{$\Pp\Pp\to\PWp\PH$} & $\sigma_{\NNLO}\, [\si{\fb}]$ \\
		\midrule
		Even-tag exclusion                     & \num{20.6828(55)}             \\
		Original flavour-$k_t$                 & \num{20.7093(63)}             \\
		\midrule
		\textbf{Ratio}                         & \textbf{99.87\%}              \\
		\bottomrule
	\end{tabular}
	\caption{Fiducial cross sections for $\PWp\PH$ at NNLO for both the original flavour-$k_t$ algorithm and our modified version where all even-tagged jets are excluded from the list of $\Pqb$-jets.
		The values are shown only for the central scales and their error represents the statistical uncertainty of the Monte Carlo integrations.}
	\label{tab:even_vs_vanilla}
\end{table}

\begin{figure}
	\centering
	\includegraphics[width=0.7\textwidth]{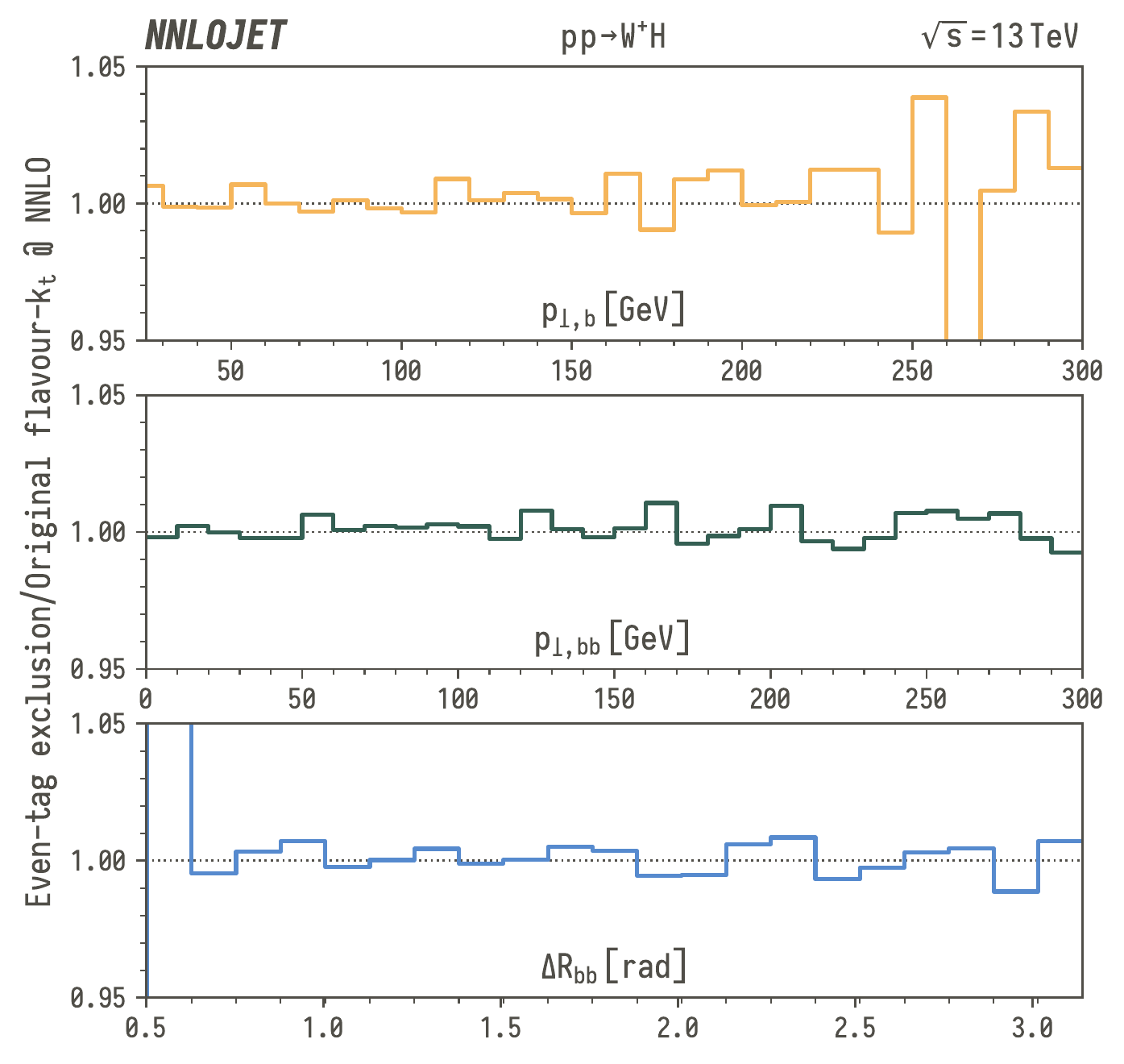}
	\caption{Bin-by-bin ratio between distributions that were calculated with the even-tag-excluded and the original variants of the flavour-$k_t$ algorithm for three observables of $\PWp\PH$: the $p_{\bot,\Pqb}$, $p_{\bot,\Pqb\Pqb}$, and $\Delta R_{\Pqb\Pqb}$ distributions at NNLO for central scale values.}
	\label{fig:even_vs_vanilla}
\end{figure}

As discussed in section~\ref{sec:jet}, the clustering outcome of the algorithm can be altered according to the criterion used to define the flavour of (pseudo)-jets. Our results have been presented with the criterion that the flavour of (pseudo)-jets is assigned as the net flavour of its constituents modulo two, which we believe is more motivated from an experimental point of view as discussed in section~\ref{sec:jet_vh}.

To investigate the impact of this ``Even-tag exclusion'' on the fixed-order predictions, we have re-computed the fiducial cross-section and distributions reported in section~\ref{sec:fiducial} and~\ref{sec:results_dist} without the additional ``modulo two'' criterion --- we refer to these results as ``Original flavour-$k_t$''.
This impact of the choice of this criterion is visualized in the case of the $\PWp\PH$ process in figure~\ref{fig:even_vs_vanilla} for the $p_{\bot,\Pqb}$, $p_{\bot,\Pqb\Pqb}$, and $\Delta R_{\Pqb\Pqb}$ distributions.
In that figure, the ratio of the two NNLO central values are divided bin-by-bin, demonstrating that this choice has no overall effect on the shape of these distributions. The small variation between bins can be attributed to statistical fluctuations.
This behaviour is also confirmed at the level of the fiducial cross section as reported in table~\ref{tab:even_vs_vanilla}, where the results are consistent within statistical uncertainties.
%
This supports our claim that no significant portion of the events are discarded by switching to the even-tag-excluded version of flavour-$k_t$ in our fixed-order predictions.

%% file: VH-AppendixB.tex
\section{Comparison with previous formulations}
\label{sec:fixbr}

As mentioned in section~\ref{sec:genframe}, NNLO-accurate observables for associated Higgs production have also been presented in~\cite{Campbell:2016jau,Ferrera:2017zex,Caola:2017xuq}.
However, the cross section in these calculations is assembled in a different manner compared with our expression in eq.~\eqref{eq:VH2}.
Specifically, the Higgs decay at the different orders had been scaled up to a fixed value of the accurately known branching ratio of the $\PH\to\Pqb\Paqb$ process.
Up to NNLO, the cross sections in this formulation is assembled as follows:
\begin{align}
	\label{eq:VH_br1}
	\rd \sigma_{\LO}^{\text{scaled}}
	 & = \rd \sigma_{V\PH}^{(0)} \times \left( \rd \sigma^{(0)}_{\PH\to\Pqb\Paqb} \right) \times K^{(0)},                                                                                     \\[10pt]
	\label{eq:VH_br2}
	\rd \sigma_{\NLO}^{\text{scaled}}
	 & = \rd \sigma_{V\PH}^{(0)} \times \left( \rd \sigma^{(0)}_{\PH\to\Pqb\Paqb} + \rd \sigma^{(1)}_{\PH\to\Pqb\Paqb} \right) \times K^{(1)} \nonumber                                       \\
	 & + \rd \sigma_{V\PH}^{(1)} \times \left( \rd \sigma^{(0)}_{\PH\to\Pqb\Paqb} \right) \times K^{(0)},                                                                                     \\[10pt]
	\label{eq:VH_br3}
	\rd \sigma_{\NNLO}^{\text{scaled}}
	 & = \rd \sigma_{V\PH}^{(0)} \times \left( \rd \sigma^{(0)}_{\PH\to\Pqb\Paqb} + \rd \sigma^{(1)}_{\PH\to\Pqb\Paqb}  + \rd \sigma^{(2)}_{\PH\to\Pqb\Paqb} \right) \times K^{(2)} \nonumber \\
	 & + \rd \sigma_{V\PH}^{(1)} \times \left( \rd \sigma^{(0)}_{\PH\to\Pqb\Paqb} + \rd \sigma^{(1)}_{\PH\to\Pqb\Paqb} \right) \times K^{(1)} \nonumber                                       \\
	 & + \rd \sigma_{V\PH}^{(2)} \times \left( \rd \sigma^{(0)}_{\PH\to\Pqb\Paqb} \right) \times K^{(0)} .
\end{align}
Here, the scaling factors $K^{(i)}$ contain the branching ratio and are given by
\begin{equation}
	K^{(i)} = \frac{\text{Br}(\PH\to\Pqb\Paqb) \, \Gamma_{\PH}}{\sum_{j=0}^{i}  \Gamma_{\PH\to\Pqb\Paqb}^{(j)}} .
\end{equation}
The branching ratio $\text{Br}(\PH\to\Pqb\Paqb)$ is kept fixed and is not a subject to an $\alphas$ expansion.

In the following, we elaborate on possible drawbacks that this prescription entails, in particular concerning theory uncertainties as estimated through scale variations.

Firstly, the scaling factors effectively divide out the Yukawa coupling $\overline{y}_{\Pqb}(\muR^{\text{dec.}}) \propto \overline{m}_{\Pqb}(\muR^{\text{dec.}})$ from the prediction.
As a result, any running of the mass as induced by the \MSbar scheme exactly cancel in the final result.
This can lead to underestimating the uncertainties, which is especially apparent at LO where the scale dependence of the Yukawa coupling otherwise dominates the uncertainties.

Secondly, analysing the structure of the scaled cross sections at NLO~\eqref{eq:VH_br2} and NNLO~\eqref{eq:VH_br3}, it is apparent that they are assembled as a sum of terms where different scaling factors $K^{(i)}$ accompany the different perturbative coefficients of the production cross section $\rd\sigma_{V\PH}^{(j)}$.
This mismatch can interfere with the compensation mechanism between terms of different orders, possibly distorting the theory error estimate obtained through variations of the production scale $\muR^{\text{prod.}}$.

\begin{table}
	\centering
	\begin{tabular}{r @{\qquad} c c c}
		\toprule
		                                                     & $\mathrm{W^+H} $                            & $\mathrm{W^-H} $                            & $ \mathrm{ZH} $                            \\
		\midrule
		$\sigma_{\text{LO}}^{\text{scaled}} \, [\si{\fb}]$   & $\num{22.52}\,^{+\num{0.63}}_{-\num{0.80}}$ & $\num{14.91}\,^{+\num{0.42}}_{-\num{0.54}}$ & $\num{6.02}\,^{+\num{0.17}}_{-\num{0.21}}$ \\
		$\sigma_{\text{NLO}}^{\text{scaled}} \, [\si{\fb}]$  & $\num{22.87}\,^{+\num{0.76}}_{-\num{0.87}}$ & $\num{15.11}\,^{+\num{0.51}}_{-\num{0.58}}$ & $\num{6.06}\,^{+\num{0.20}}_{-\num{0.23}}$ \\
		$\sigma_{\text{NNLO}}^{\text{scaled}} \, [\si{\fb}]$ & $\num{20.93}\,^{+\num{0.61}}_{-\num{0.73}}$ & $\num{13.80}\,^{+\num{0.41}}_{-\num{0.49}}$ & $\num{6.10}\,^{+\num{0.31}}_{-\num{0.31}}$ \\
		\bottomrule
	\end{tabular}
	\caption{The scaled fiducial cross sections for all $V\PH$ processes according to the setup of appendix~\ref{sec:fixbr} at each perturbative order up to $\cO(\alphas^2)$.}
	\label{tab:br}
\end{table}

To quantify the differences between the two approaches, in table~\ref{tab:br} we report the fiducial cross sections obtained according to~\eqref{eq:VH_br1}--\eqref{eq:VH_br3} using $\mathrm{Br}(\PH\to\Pqb\Paqb) = \num{58.09}\%$~\cite{deFlorian:2016spz}.
Comparing these predictions with those given in table~\ref{tab:fid} using the unscaled cross section formul\ae~\eqref{eq:VH1}, we observe that the central value of the LO prediction is substantially improved in the scaled predictions thanks to absorbing higher-order effects to the $\PH\to\Pqb\Paqb$ decay through the branching ratio.
At NLO, however, the scaled prediction appears to slightly overestimate the cross section, while the associated theory uncertainties are comparable in size between the two formulations.
At NNLO, both prescriptions agree well in their respective central values, however, sizeable differences can be seen in their associated uncertainties.
The scaled predictions at NNLO show almost no reduction in scale uncertainties --- even increasing for $\PZ\PH$ --- compared to the respective NLO number, whereas our formulation~\eqref{eq:VH1} exhibits a substantial reduction in scale uncertainties when going from NLO to NNLO.
This difference can be attributed to the aforementioned compensation of scale dependences, which is spoiled by the different rescaling factors used in eq.~\eqref{eq:VH_br3}.

The effects of dividing out the Yukawa coupling in the decay and the scaling factor mismatch during the assembly of production cross sections are apparent as the theoretical uncertainties of the NNLO cross section barely change compared to their NLO values.
In our opinion, the consistent treatment of theoretical uncertainties outweighs the precision gain that one might (or might not) get by scaling to a fixed branching ratio, especially in the case of NNLO-accurate observables.
This further motivates our initial and simpler formulation we presented in eq.~\eqref{eq:VH1} of section~\ref{sec:genframe} where no scaling factors are applied.